\documentclass[a4]{article}

\usepackage[margin=2cm]{geometry}

\usepackage[none]{hyphenat}

\usepackage{url}
\usepackage{graphicx}

\begin{document}

\title{Collective Tuning Initiative: automating and accelerating development and optimization of computing systems}

\date{June 14, 2009}             % You can put a fixed date in if you wish,
                    % allow LaTeX to use the date of typesetting,
                    % or use \date{} to have no date at all.
                    % Whatever you do, there will not be a date
                    % shown in the proceedings.

\author{Grigori Fursin \\
{\itshape INRIA, France}\\
{\itshape HiPEAC member}\\[.3cm]
{\ttfamily\normalsize grigori.fursin@inria.fr}\\
} % end author section

%%%%%%%%%%%%%%%%%%%%%%%%%%%%%%%%%%%%%%%%%%%%%%%%%%%%%%%%%%%%%%%%%%%%%%%%%
\maketitle

\centerline{\textbf{\textit{Proceedings of the GCC Developers' Summit'09, 14 June 2009, Montreal, Canada}}}

\begin{abstract}
\small

Computing systems rarely deliver best possible performance due to
ever increasing hardware and software complexity and limitations
of the current optimization technology. Additional code and 
architecture optimizations are often required to improve execution time, 
size, power consumption, reliability and other important characteristics
of computing systems. However, it is often a tedious, repetitive, isolated 
and time consuming process. In order to automate, simplify and systematize
program optimization and architecture design, we are developing
open-source modular plugin-based Collective Tuning Infrastructure (http://cTuning.org) 
that can distribute optimization process and leverage optimization experience 
of multiple users.

The core of this infrastructure is a Collective Optimization Database that
allows easy collection, sharing, characterization and reuse of a large
number of optimization cases from the community. The infrastructure also
includes collaborative R\&D tools with common API (Continuous Collective
Compilation Framework, MILEPOST GCC with Interactive Compilation Interface
and static feature extractor, Collective Benchmark and Universal Run-time
Adaptation Framework) to automate optimization, produce adaptive
applications and enable realistic benchmarking. We developed several tools
and open web-services to substitute default compiler optimization heuristic
and predict good optimizations for a given program, dataset and architecture
based on static and dynamic program features and standard machine learning
techniques.

Collective tuning infrastructure provides a novel fully integrated,
collaborative, "one button" approach to improve existing underperfoming
computing systems ranging from embedded architectures to high-performance
servers based on systematic iterative compilation, statistical collective
optimization and machine learning. Our experimental results show that it is
possible to reduce execution time (and code size) of some programs from
SPEC2006 and EEMBC among others by more than a factor of 2 automatically. It
can also reduce development and testing time considerably.
Together with the first production quality machine learning enabled
interactive research compiler (MILEPOST GCC) this infrastructure opens up
many research opportunities to study and develop future realistic
self-tuning and self-organizing adaptive intelligent computing systems based
on systematic statistical performance evaluation and benchmarking. Finally,
using common optimization repository is intended to improve the quality and
reproducibility of the research on architecture and code optimization.

\end{abstract}

\textbf{Keywords}
\small{
\textit{
systematic benchmarking, performance analysis, performance tuning, 
energy minimization, size minimization, automatic program optimization,
architecture design, software and hardware co-design,
plugin-based auto-tuning, optimization space exploration,
adaptive sampling, predictive analytics, machine learning, data
mining, statistical analysis, feature detection, big data,
crowdsourcing, collective intelligence, machine learning based
compilation, MILEPOST GCC, plugin interface for compilers,
run-time adaptation, dynamic adaptation for static binaries, static code
multi-versioning, reproducible research, artifact sharing, common
experimental setups, collaborative artifact evaluation, community
driven research, reproducible research, open publication model
}

%%%%%%%%%%%%%%%%%%%%%%%%%%%%%%%%%%%%%%%%%%%%%%%%%%%%%%%%%%%%%%%%%%%%%%%%%
\section{Introduction}

\begin{figure}[htb]
 \centering
 \includegraphics[width=6in]{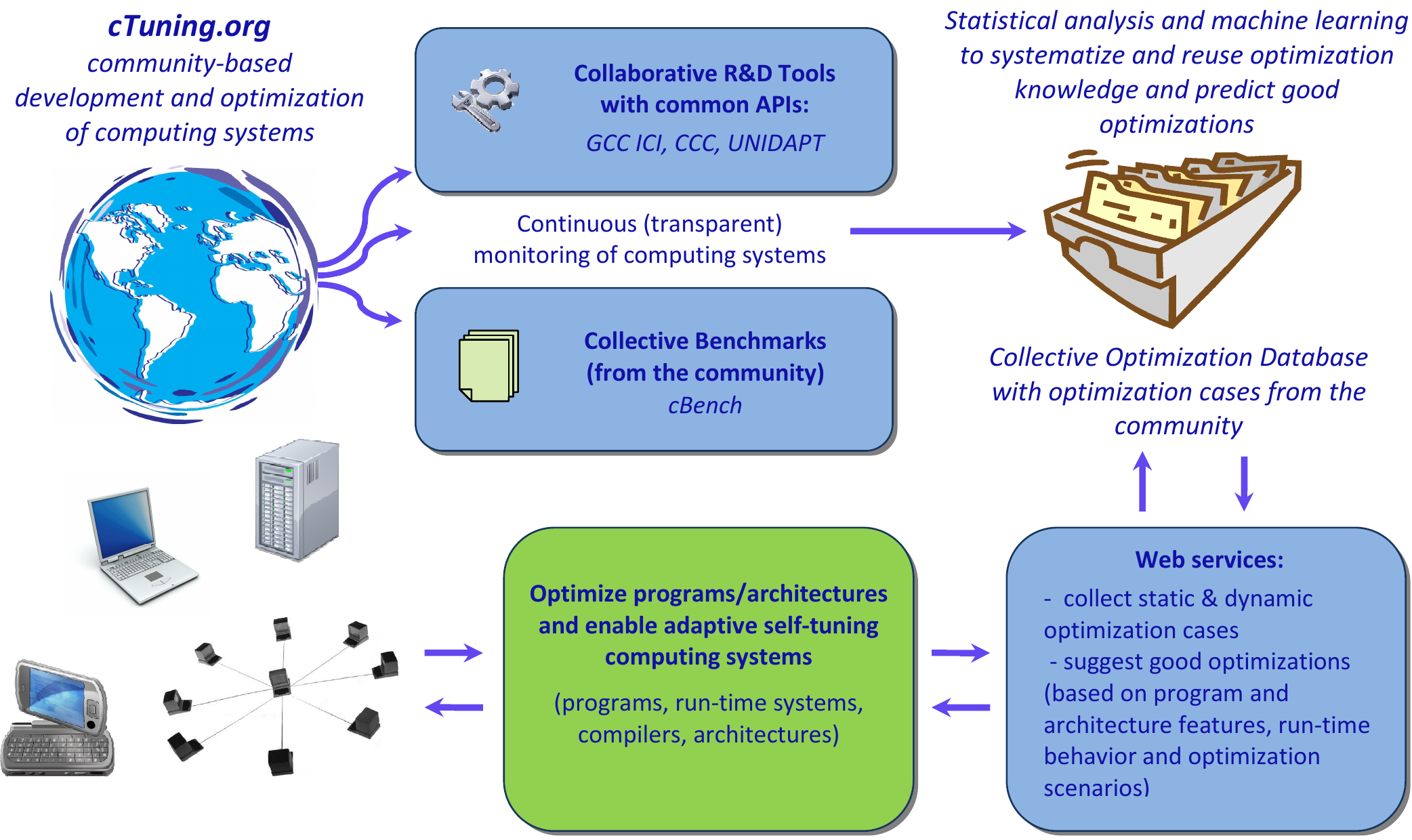}
 \caption{Collective tuning infrastructure to enable systematic collection, sharing and reuse 
 of optimization knowledge from the community. It automates optimization of computing systems
 by leveraging the optimization experience from multiple users.}
 \label{fig:ctuning}
\end{figure}

Continuing innovation in science and technology requires increasing
computing resources while imposing strict requirements on cost, 
performance, power consumption, size, response time, 
reliability, portability and design time of computing systems. 
Embedded and large-scale systems tend to evolve towards complex 
heterogeneous reconfigurable multiprocessing systems with 
dramatically increased design, test and optimization time.

A compiler is one of the key components of computing systems 
responsible for delivering high-quality machine code across a wide range
of architectures and programs with multiple inputs. However, for several 
decades compilers fail to deliver portable performance
often due to restrictions on optimization time, simplified cost models
for rapidly evolving complex architectures, large number of combinations 
of available optimizations, limitations on run-time adaptation and inability to 
leverage optimization experience from multiple users efficiently, systematically 
and automatically. 

Tuning compiler optimization heuristic for a given architecture is a repetitive
and time consuming process because of a large number of possible transformations, 
architecture configurations, programs and inputs available as well as multiple 
optimization objectives such as improving performance, code size, reliability among others.
Therefore, when adding new optimizations or retargeting to a new architecture, compilers are 
often tuned only for a limited set of architecture configurations, benchmarks and transformations
thus making even relatively recent computing systems underperform. 
Hence, most of the time, users have to resort to additional optimizations 
to improve the utilization of available resources of their systems.

Iterative compilation has been introduced to automate program optimization
for a given architecture using an empirical feedback-directed search for good program
transformations~\cite{atlas,CSS99,pfdc,fftw,FOK02,Fur2004,CST02,acovea,suif_compiler,rose_compiler,esto,pathscale,vista,TVVA03,spiral,PE2006,HB2006,HE2008}.
Recently, the search time has been considerably reduced using statistical
techniques, machine learning and continuous optimization~\cite{Monsifrot,metaOpt03,SA2005,soffa2005,VE00,LCYP04,la2004,ABCP06,FMTP2008}.
However, iterative feedback-directed compilation is often performed with the
same program input and has to be repeated if dataset changes. In order to
overcome this problem, a framework has been developed to statically enable
run-time optimizations based on static function multiversioning, iterative
compilation and low-overhead run-time program behavior monitoring
routines~\cite{FCOP2005,LCWP2009} (a similar framework has also been presented
in~\cite{MH2009} recently).

Though these techniques demonstrated significant performance improvements,
they have not yet been fully adopted in production environments due to a
large number of training runs required to test many different combinations
of optimizations. Therefore, in~\cite{FT2009} we proposed to overcome this
obstacle and speed up iterative compilation using statistical collective
optimization, where the task of optimizing a program leverages the
experience of many users, rather than being performed in isolation, and
often redundantly, by each user. To some extend it is similar to biological
adaptive systems since all programs for all users can be randomly modified
(keeping the same semantics) to explore some part of large optimization
spaces in a distributed manner and favor the best performing optimizations
to improve computing systems continuously.

Collective optimization requires radical changes to the current compiler and architecture design
and optimization technology. In this paper we present a long term community-driven collective tuning initiative 
to enable collective optimization of computing systems based on systematic, automatic and distributed exploration 
of program and architecture optimizations, statistical analysis and machine learning. 
It is based on a novel fully integrated collaborative infrastructure with common optimization repository 
(Collective Optimization Database) and collaborative R\&D tools with common APIs (including
the first of its kind production quality machine learning enabled research
compiler (MILEPOST GCC)~\cite{FMTP2008}) to share profitable optimization 
cases and leverage optimization experience from multiple users automatically.

We decided to use a top-down systematic optimization approach providing capabilities 
for global and coarse-grain optimization, parallelization and run-time adaptation first,
and then combining it with finer grain optimizations at a loop or instruction level.
We believe that this is the right approach to avoid the tendency to target
very fine grain optimizations at first without solving the global optimization problem
that may have much higher potential benefits. 
     
A collective tuning infrastructure can already improve a broad range of existing 
desktop, server and embedded computing systems using empirical iterative compilation
and machine learning techniques. We managed to reduce the execution time
(and code size) of multiple programs from SPEC95,2000,2006, EEMBC v1 and v2, cBench
ranging from several percent to more than a factor of 2 on several common x86 architectures.
On average, we reduced the execution time of the cBench benchmark suite
for ARC725D embedded reconfigurable processor by 11\% entirely automatically. 

Collective tuning technology helps to minimize repetitive time consuming tasks and human
intervention and opens up many research opportunities. Such community-driven collective 
optimization technology is the first step towards our long term objective to study and 
develop smart self-tuning adaptive heterogeneous multi-core computing systems. 
We also believe that our initiative can improve the quality and reproducibility
of academic and industrial IT research on code and architecture design and optimization.
Currently, it is not always easy to reproduce and verify experimental results of multiple 
research papers that should not be acceptable anymore. Using common optimization repository 
and collaborative R\&D tools provides means for fair and open comparison of available 
optimization techniques, helps to avoid overstatements and mistakes, and should eventually
boost innovation and research. 

The paper is organized as follows. Section~\ref{sec:ctc} introduces a collective tuning infrastructure.
Section~\ref{sec:rep} presents a collective optimization repository to share and leverage optimization experience
from the community. Section~\ref{sec:tools} presents collaborative R\&D tools and cBench 
to automate, systematize and distribute optimization exploration.
Finally, section~\ref{sec:scenarios} provides some practical usage scenarios followed
by the future research and development directions.
             
%%%%%%%%%%%%%%%%%%%%%%%%%%%%%%%%%%%%%%%%%%%%%%%%%%%%%%%%%%%%%%%%%%%%%%%%%

\section{Collective Tuning Infrastructure}
\label{sec:ctc}

In order to enable systematic and automatic collection, sharing and reuse of
profiling and optimization information from the community, we develop a
fully integrated collective tuning infrastructure shown in
Figure~\ref{fig:ctuning}. The core of the cTuning infrastructure is an
extendable optimization repository (Collective Optimization Database) to
characterize multiple heterogeneous optimization cases from the community 
which improve execution time and code size, track program and compiler bugs 
among many others and to ensure their reproducibility. It is described in detail in
Section~\ref{sec:rep}. Optimization data can be searched and analyzed using
web services with open APIs and external plugins. A user can submit
optimization cases either manually using an online submission form
at~\cite{ctuning_repository} or automatically using collaborative R\&D
tools (cTools)~\cite{ctuning_tools} described in Section~\ref{sec:tools}.

The current cTools include:

\begin{itemize}
\item Extensible plugin-enabled Continuous Collective Compilation Framework
(CCC) to automate empirical iterative feedback-directed compilation and
allow users to explore a part of large optimizations spaces in a distributed
manner using multiple search strategies.
\item Extensible plugin-enabled GCC with high-level Interactive Compilation Interface (ICI)
to open up production compilers and control their compilation flow and decisions using
external user-defined plugins. Currently, it allows selection and tuning of compiler optimizations 
(global compiler flags, passes at function level and fine-grain transformations) as well as program analysis
and instrumentation.
\item Open-source plugin-based machine learning enabled interactive research
compiler based on GCC (MILEPOST GCC)~\cite{FMTP2008} that includes ICI and a static program
feature extractor to substitute default optimization heuristic of a compiler
and predict good optimizations based on static and dynamic program features 
(some general aspects of a program) and machine learning.
\item Universal run-time adaptation framework (UNIDAPT) to enable
transparent monitoring of dynamic program behavior as well as dynamic
optimization and adaptation if statically compiled programs with multiple datasets
for uni-core and heterogeneous reconfigurable multi-core architecture based on code
multiversioning.
\end{itemize}

An automatic exploration of optimization spaces is performed using multiple
publicly-available realistic programs and their datasets from the community that
compose cBench~\cite{ctuning_cbench}. However, we also plan to use 
our collective optimization approach when stable to enable a fully transparent collection
of optimization cases from multiple users~\cite{FT2009}. 

\begin{figure}[htb]
 \centering
 \includegraphics[width=5in]{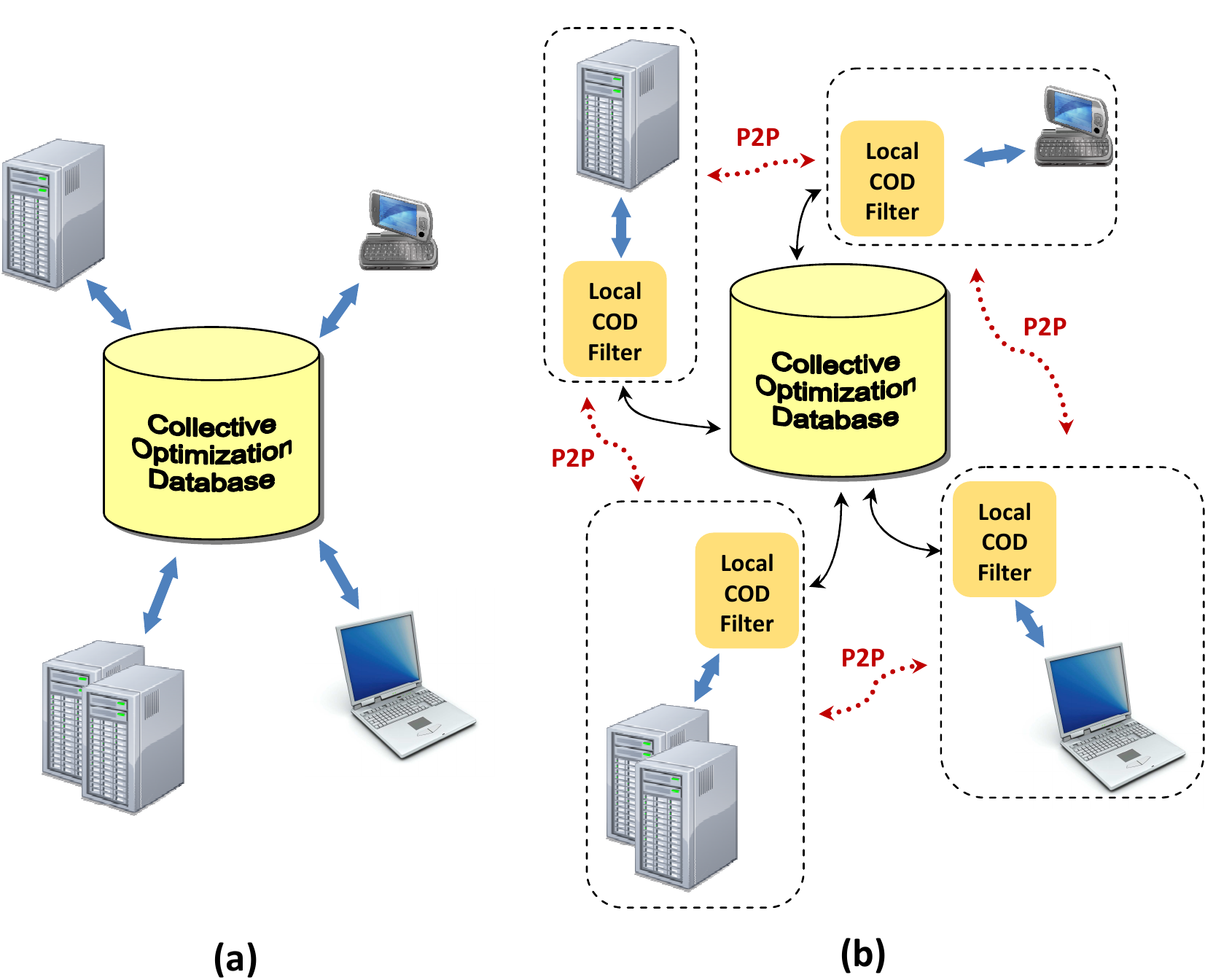}
 \caption{(a) original centralized design of COD with large communication overheads (b) decentralized design of COD with local filters to prune optimization information and minimize communication costs.}
 \label{fig:cod_design}
\end{figure}

The cTuning infrastructure is available online at the community-driven
wiki-based web portal~\cite{ctuning} and has been extended within MILEPOST
project~\cite{milepost}. It now includes plugins to automate compiler and
architecture design, substitute default GCC optimization heuristic and
predict good program optimizations for a wide range of architectures using
machine learning and statistical collective optimization
plugins~\cite{FT2009,FMTP2008}. But more importantly, it creates a common platform 
for innovation and opens up multiple research possibilities for the academic 
community and industry.

%%%%%%%%%%%%%%%%%%%%%%%%%%%%%%%%%%%%%%%%%%%%%%%%%%%%%%%%%%%%%%%%%%%%%%%%%
\section{Collective Optimization Database}
\label{sec:rep}

Collective Optimization Database (COD) is the key component of the cTuning infrastructure serving
as a common extensible open online repository of a large number of optimization cases from the community. 
Such cases include program optimizations and architecture configurations to improve
execution time, code size, power consumption or detect performance anomalies and bugs, etc.
COD should be able to keep enough information to describe optimization cases and
characterize program compilation and optimization flow, run-time behavior and architecture parameters 
to ensure reproducibility and portable performance for collective optimization.

Before the MILEPOST project, COD had a fully centralized design shown in Figure~\ref{fig:cod_design}a with
all the data coming directly from users. Such design may cause large communication overheads
and database overloading thus requiring continuous resource-hungry pruning of the data on the database server.
Therefore, the design of COD  has been gradually altered to support 
local user-side filtering of optimization information using plugins of CCC framework
as shown in Figure~\ref{fig:cod_design}b. Currently, plugin-based filters detect
optimization cases that improve execution time and code size based on Pareto-like distributions
or that has some performance anomalies to allow further detailed analysis. We gradually extend 
these filters to detect important program and architecture optimizations as well as useful static
and dynamic program features automatically based on Principle Component Analysis and similar techniques~\cite{CFAP2007}
to improve the correlation between program structure or behavior and optimizations.

We also investigate the possibility to develop a fully decentralized collective tuning system 
to enable continuous exploration, analysis and filtering of program optimizations, static/dynamic features
and architecture configurations as well as transparent sharing and reuse of optimization knowledge between 
multiple users based on P2P networks. 

\begin{figure}[htb]
 \centering
 \includegraphics[width=6in]{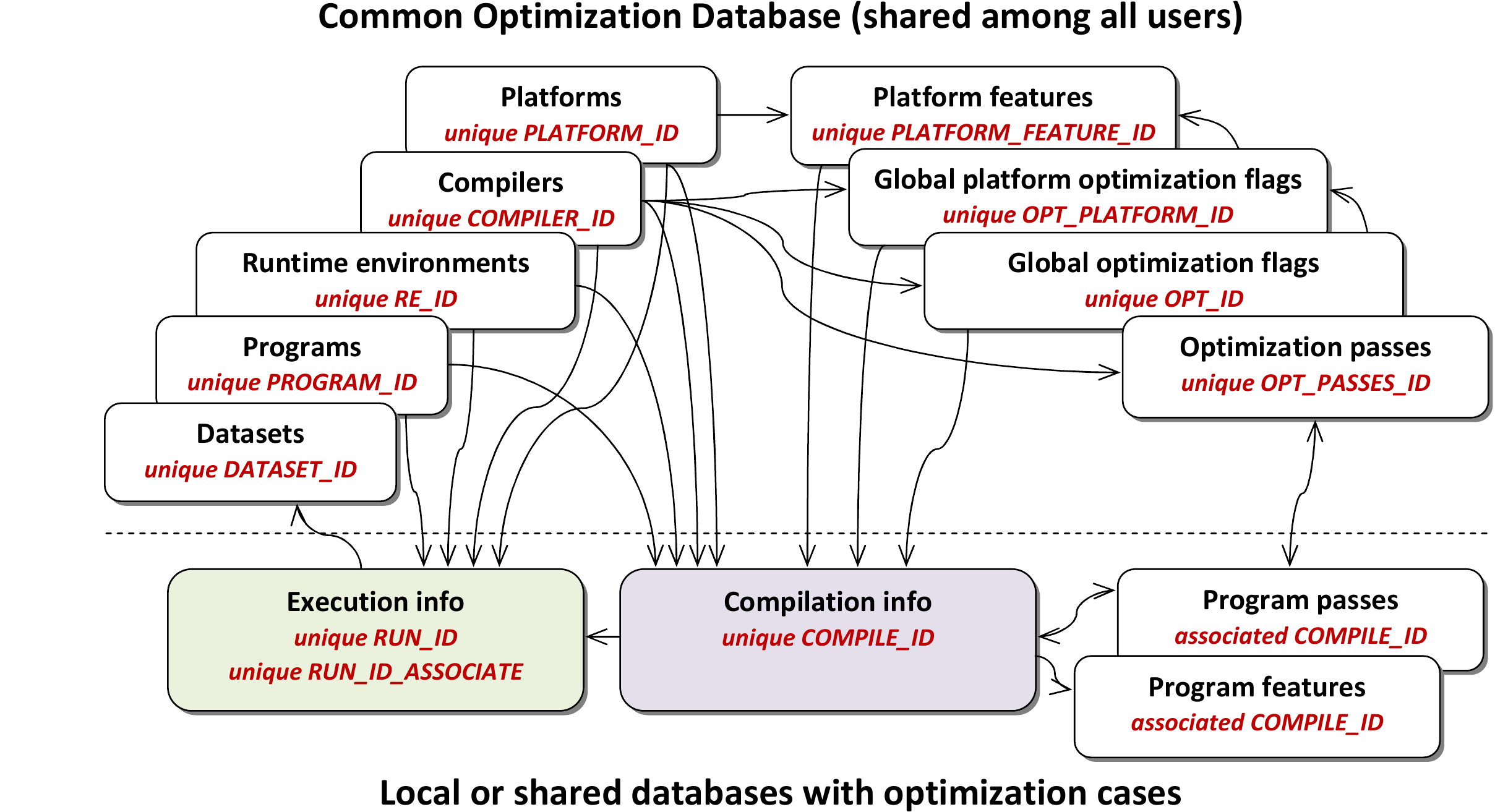}
 \caption{Collective Optimization Database structure (tables) to describe optimization cases and minimize database size:
common informative part and shared or private part with optimization cases.}
 \label{fig:cod_structure}
\end{figure}

Current design of COD presented in Figure~\ref{fig:cod_structure} has been
influenced by the requirements of the MILEPOST project~\cite{milepost} to
collect a large number of optimization cases for different programs,
datasets and architectures during iterative feedback-directed compilation
from several partners. These cases are used to train machine learning model and
predict good optimizations for a given program on a given reconfigurable
architecture based on program static or dynamic features~\cite{FMTP2008,CFAP2007}.

Before participating in collective tuning and sharing of optimization cases,
users must register their computing systems or find similar ones from
the existing records. This includes information about their computing platforms 
(architecture, GPUs, accelerators, memory and HDD parameters, etc), software environment 
(OS and libraries), compiler and run-time environment (VM or architecture simulator, if used). 
Users can participate in distributed exploration of optimization spaces using cBench
that is already prepared to work directly with the cTuning
infrastructure. Alternatively, users can register and prepare their own
programs and datasets to support cTuning infrastructure using CCC framework. 
Currently, we are extending the cTuning infrastructure and GCC
to enable transparent program optimization without Makefile or project modifications~\cite{gsoc2009}
based on collective optimization concept~\cite{FT2009} and UNIDAPT framework~\cite{ctuning_unidapt}.

All the information about computing systems is recorded in COD and shared among all users. 
All the records have unique UUID-based identifiers to enable full decentralization of the 
infrastructure and unique referencing of optimization cases by the community 
(in reports, documentation and research publications for example).

\begin{figure}[htbp]
 \centering
  {\footnotesize
  \begin{tabular}{|l|p{280pt}|}
    \hline
    \textbf{Field} & \textbf{Description:} \\
    \hline
      COMPILE\_ID                     & Unique UUID-based identifier to enable global referencing of a given optimization case \\
      PLATFORM\_ID                    & Unique platform identifier \\
      ENVIRONMENT\_ID                 & Unique software environment identifier \\
      COMPILER\_ID                    & Unique compiler identifier \\
      PROGRAM\_ID                     & Unique program identifier \\
      PLATFORM\_FEATURE\_ID           & Reference to the table with platform features describing platform specific features
for architectural design space exploration (it can include architecture specific flags such as -msse2 or cache parameters, etc)\\
      OPT\_ID                         & Reference to the table with global optimization flags \\
      COMPILE\_TIME                   & Overall compilation time\\
      BIN\_SIZE                       & Binary size\\
      OBJ\_MD5CRC                     & MD5-based CRC of the object file to detect whether optimizations changed the code or not\\
      ICI\_PASSES\_USE                & set to 1 if compiler with Interactive Compilation Interface (ICI) has been used to select
individual passes and their orders on a function level \\
      ICI\_FEATURES\_STATIC\_EXTRACT  & set to 1 if static program features has been extracted using ICI and MILEPOST GCC~\cite{FMTP2008} \\
      OPT\_FINE                       & XML description of fine-grain optimizations selected using ICI (on-going work) \\
      OPT\_PAR\_STATIC                & XML description of static program parallelization (on-going work) \\
      NOTES                           & User notes about this optimization case (can describe a bug or unusual compiler behavior for further analysis
for example)\\
    \hline
  \end{tabular}
 }
 \caption{\label{fig:table_comp} Summary of current fields of the COD table describing compilation process.}
\end{figure}

Each optimization case is represented by a combination of program compilations (with different optimizations
and architecture configurations) and executions (with the same or different dataset and 
run-time environment). The optimization information from a user is first collected in the Local Optimization Database to enable
data filtering, minimize global communication costs and provide a possibility for internal non-shared
optimizations within companies. Each record in the databases has a unique UUID-based identifier to simplify
merging of the distributed filtered data in COD. 

Information about compilation process is recorded in a special COD table described in Figure~\ref{fig:table_comp}.
In order to reduce the size of the database, the information that can be shared among multiple users or among
multiple optimization cases from a given user has been moved to special common tables. This includes global optimization
flags, architecture configuration flags (such as -msse2, -mA7, -ffixed-r16, -march=athlon64, -mtune=itanium2)
and features, sequences of program passes applied to functions when using a compiler with the Interactive Compilation Interface, 
program static features extracted using MILEPOST GCC~\cite{FMTP2008} among others.

\begin{figure}[htbp]
 \centering
 {\footnotesize
  \begin{tabular}{|l|p{280pt}|}
    \hline
    \textbf{Field} & \textbf{Description:} \\
    \hline
      RUN\_ID                & Unique UUID-based identifier to enable global referencing of a given optimization case \\
      RUN\_ID\_ASSOCIATE     & ID of the associated run with baseline optimization for further analysis\\
      COMPILE\_ID            & Associated compilation identifier \\
      COMPILER\_ID           & Duplicate compiler identifier from the compilation table to speed up SQL queries \\
      PROGRAM\_ID            & Duplicate program identifier from the compilation table to speed up SQL queries \\
      BIN\_SIZE              & Duplicate binary size from the compilation table to speed up SQL queries \\
      PLATFORM\_ID           & Unique platform identifier \\
      ENVIRONMENT\_ID        & Unique software environment identifier \\
      RE\_ID                 & Unique runtime environment identifier \\
      DATASET\_ID            & Unique dataset identifier \\
      OUTPUT\_CORRECT        & Set to 1 if the program output is the same as the reference one 
(supported when using Collective Benchmark or program specially prepared using CCC framework). It is important
to add formal validation methods in the future particularly for transparent collective optimization~\cite{FT2009}. \\
      RUN\_TIME              & Absolute execution time in seconds (or relative number if absolute time can not be disclosed
by some companies or when using some benchmarks) \\
      RUN\_TIME\_USER        & User execution time \\
      RUN\_TIME\_SYS         & System execution time \\
      RUN\_TIME\_BACKGROUND  & Information about background processes to be able to analyze the interference 
between multiple running applications and enable better adaptation and scheduling when sharing resources on uni-core 
or multi-core systems~\cite{JGVP2009}\\
      RUN\_PG                & Function-level profile information (using gprof or oprofile): $\langle$function name={time spent in this function}, ...$\rangle$ \\
      RUN\_HC                & Dynamic program features (using hardware counters): $\langle$hardware counter=value, ...$\rangle$ \\
      RUN\_POWER             & Power consumption (on-going work) \\
      RUN\_ENERGY            & Energy during overall program execution (on-going work) \\
      PAR\_DYNAMIC           & Information about dynamic dependencies to enable dynamic parallelization (on-going work) \\
      PROCESSOR\_NUM         & Core number assigned to the executed process \\
      RANK                   & Integer number describing ranking (profitability) of the optimization. 
Optimization case can be ranked manually  or automatically based on statistical collective optimization~\cite{FT2009} \\
      NOTES                  & User notes about this optimization case\\\\
    \hline
  \end{tabular}
 }
 \caption{\label{fig:table_run} Summary of current fields of the COD table to describe program executions.}
\end{figure}

Information about program execution is recorded in a special COD table
described in Figure~\ref{fig:table_run}. Though an absolute execution time can
be important for benchmarking and other reasons, we are more interested
in how optimization cases improve execution time, code size or other
characteristics. In the case of "traditional" feedback-directed compilation, 
we need two or more runs with the same dataset to evaluate the impact of
optimizations on execution time or other metrics: one with the reference
optimization level such as -O3 (referenced by RUN\_ID\_ASSOCIATE) and
another with a new set of optimizations and exactly the same dataset
(referenced by RUN\_ID). When a user explores larger optimization space using
CCC framework for a given program with a given dataset, the obtained
combination of multiple optimization cases includes the same associated reference
id (RUN\_ID\_ASSOCIATE) to be able to calculate improvements over
original execution time or other metrics. We perform multiple runs with the same
optimization and the same dataset to calculate speedup confidence and deal 
with timer/hardware counters noise. We use the MD5 CRC of the executable 
(OBJ\_MD5CRC) to compare a transformed code with the original one and avoid 
executing code when optimizations did not affect the code.

When sharing multiple optimization cases among users, there is a natural competition between different
optimizations that can improve a computing system. Hence, we use a simple ranking system to favor stable 
optimization cases across the largest number of users. Currently, users rank optimization cases manually,
however we plan to automate this process using statistical ranking of optimizations described in~\cite{FT2009}.
This will require extensions to the UNIDAPT framework~\cite{ctuning_unidapt} described later 
to enable transparent evaluation of program optimizations with any dataset during single execution 
without a need for a reference run based on static multiversioning and statistical run-time optimization 
evaluation~\cite{FCOP2005,FT2009,ctuning_tools}. 

At the moment, COD uses MYSQL engine and can be accessed either directly or
through online web-services. The full description of the COD structure and
web-service is available at the collective tuning
website~\cite{ctuning_repository}. Since collective tuning is on-going
long term initiative, the COD structure may evolve over time. Hence, we
provide current COD version number in the INFORMATION table to ensure compatibility
between all cTuning tools and plugins that access COD.

%%%%%%%%%%%%%%%%%%%%%%%%%%%%%%%%%%%%%%%%%%%%%%%%%%%%%%%%%%%%%%%%%%%%%%%%%
\section{Collaborative R\&D Tools}
\label{sec:tools}

Many R\&D tools have been developed in the past few decades to enable
empirical iterative feedback-directed optimization and analysis
including~\cite{acovea,suif_compiler,rose_compiler,atlas,esto,pathscale,spiral,vista}.
However they are often slow, often domain, compiler and platform specific, and are not
capable of sharing and reusing optimization information about different
programs, datasets, compilers and architectures.
Moreover, they are often not compatible with each other, not fully supported, 
are unstable and sometimes do not provide open sources to enable further extensions. 
As a result, iterative feedback-directed compilation has not yet been widely adopted.

Previously, we have shown the possibility for realistic, fast and automatic
program optimization and compiler/architecture co-tuning based on empirical
optimization space exploration, statistical analysis, machine learning and
run-time adaptation~\cite{FOK02,Fur2004,ABCP06,FCOP2005,FOTP2005,CFAP2007,FMTP2008,LCWP2009,FT2009}.
After we obtained promising results and our techniques become more mature,
we decided to initiate a rigorous systematic evaluation and
validation of iterative feedback-directed optimization techniques across
multiple programs, datasets, architectures and compilers but faced a lack of
generic, stable, extensible and portable open-source infrastructure to support this
empirical study with multiple optimization search strategies. 
Hence, we decided to develop and connect all our tools, benchmarks and databases together 
within Collective Tuning Infrastructure using open APIs and move all our developments 
to the public domain~\cite{ctuning_tools,ctuning_mailing_lists} to extend our techniques
and enable further collaborative and systematic community-driven R\&D to automate code and architecture
optimization and enable future self-tuning adaptive computing systems.

Instead of developing our own source-to-source transformation and instrumentation
tools, we decided to reuse and "open up" existing production quality compilers
using an event-driven plugin system called Interactive Compilation Interface (ICI)~\cite{ctuning_ici,FCOP2005,FC2007,FMTP2008}. 
We decided to use GCC for our project since it is a unique open-source production quality compiler 
that supports multiple architectures, languages and has a large user base. 
Using a plugin-enabled production compiler can improve the quality and reproducibility 
of research and help to move research prototypes back to a compiler much faster for the benefit
of the whole community. 

Finally, we are developing a universal run-time adaptation framework
(UNIDAPT) to enable transparent collective optimization, run-time adaptation and split
compilation for statically compiled programs with multiple datasets across
different uni-core and multi-core heterogeneous architectures and
environments~\cite{FCOP2005,LCWP2009,FT2009,JGVP2009}. We have also collected
multiple datasets within Collective Benchmark (formerly MiBench/MiDataSets)
to enable realistic research on program optimization with multiple inputs.

We hope that using common tools will help to avoid costly duplicate developments, will improve
quality and reproducibility of the research and will boost innovation in program optimization, 
compiler design and architecture tuning. 

%%%%%%%%%%%%%%%%%%%%%%%%%%%%%%%%%%%%%%%%%%%%%%%%%%%%%%%%%%%%%%%%%%%%%%%%%%%%%%%%%%%%%%
\subsection{Continuous Collective Compilation Framework}

In~\cite{FOK02,FOTP04,Fur2004} we demonstrated the possibility to apply iterative feedback-directed
compilation to large applications at a loop level. It was a continuation of the MHAOTEU project 
(1999-2000)~\cite{ATAP2000} where we had to develop a source-to-source Fortran 77 and C compiler to enable 
parametric transformations such as loop unrolling, tiling, interchange, fusion/fission, array padding and some
others, evaluate their effect on large and resource-hungry programs, improve their memory utilization
and execution time on high-performance servers and supercomputers, and predict the performance upper bound
to guide iterative search. The MHAOTEU project was in turn a continuation of the 
OCEANS project (1996-1999) where general iterative feedback-directed compilation was introduced
to optimize relatively small kernels and applications for embedded computing systems~\cite{europar97}.

\begin{figure}[htb]
 \centering
 \includegraphics[width=6in]{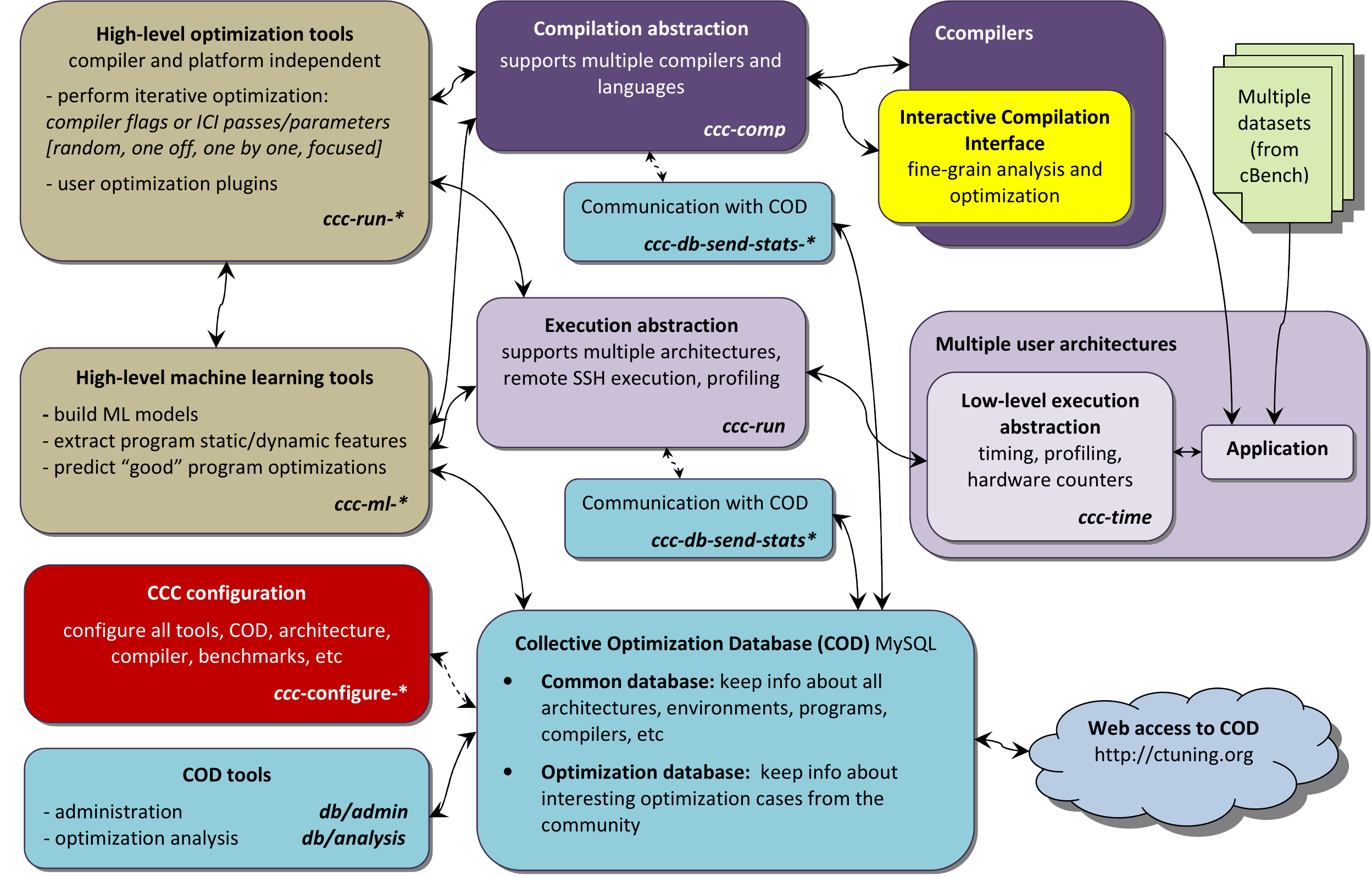}
 \caption{CCC framework enables systematic, automatic and distributed program optimization, 
compiler design and architecture tuning with multiple optimization strategies 
as well as collection of optimization and profiling statistics from multiple users in COD
for further analysis and reuse.}
 \label{fig:ccc}
\end{figure}

In order to initiate further rigorous systematic evaluation and validation of
iterative code and architecture optimizations, we started developing  our
own open-source modular Continuous Collective Compilation framework (CCC).
CCC framework is intended to automate program optimization, compiler design
and architecture tuning using empirical iterative feedback-directed
compilation. It enables collaborative and distributed exploration of program
and architecture optimization spaces and collects static and dynamic
optimization and profile statistics in COD. CCC has been designed using
modular approach as shown in figure~\ref{fig:ccc}. It has several low-level
platform-dependent tools and platform-independent tools to abstract
compilation and execution of programs. It also includes routines to
communicate with COD and high-level plugins for iterative compilation with
multiple search strategies, statistical analysis and machine learning.

CCC is installed using INSTALL.sh script from the root directory. During installation, several scripts
are invoked to configure the system, provide info about architectures, environments, compilers, runtime systems and set up
an environment. Before compilation of platform-dependent tools, a user must select the how to store optimization
and profile information, i.e. within local or shared COD at each step,
or as a minimal off-line mode when all statistics is recorded
in a local directory of each optimized program. An off-line mode can be useful for GRID-like environments with 
network filters such as GRID5000~\cite{grid5000} where all statistics from multiple experiments can be 
aggregated in a text file and recorded in COD later. CCC may require PHP,
PAPI library, PapiEx, uuidgen and oprofile for extended functionality and MYSQL client
to work with local and shared COD (Figure~\ref{fig:cod_structure}) configured 
using the following environment variables:

\begin{itemize}
\item \emph{CCC\_C\_URL, CCC\_C\_DB, CCC\_C\_USER, CCC\_C\_PASS, CCC\_C\_SSL} for a common database (URL or IP of the server, database name, username, password and SSL attributes for secure access
\item \emph{CCC\_URL, CCC\_DB, CCC\_USER, CCC\_PASS, CCC\_SSL} for a local database with optimization cases
\item \emph{CCC\_CT\_URL, CCC\_CT\_DB, CCC\_CT\_USER, CCC\_CT\_PASS, CCC\_CT\_SSL} for a shared database with filtered optimization cases visible at~\cite{ctuning_repository}
\end{itemize}

We decided to utilize a systematic top-down approach for optimizations: though
originally we started our research on iterative compilation from fine-grain
loop transformations~\cite{ATAP2000,FOK02,Fur2004}, we think that the current research on
architecture and code optimizations focuses too much on solving
only local issues while sometimes overlooking global and coarse-grain problems. 
Therefore, we first developed optimization space exploration plugins with various search
algorithms for global compiler flags and started gradually adding support
for finer grain program optimizations using plugin-enabled GCC with ICI
(described in Section~\ref{sec:tools:ici}).

Originally, we started using Open64/PathScale compilers for our systematic empirical
studies~\cite{FCOP2005,CDAP2006,FCOP2007} but gradually moved to GCC since
it is the only open-source production-quality compiler that supports
multiple architectures (more than 30 families) and languages. It also
features many aggressive optimizations including a unique GRAPHITE
coarse-grain polyhedral optimization framework. However, CCC framework can
be easily configured to support multiple compilers including
LLVM,GCC4NET,Open64,Intel ICC,IBM XL and others to enable a fair comparison of
different compilers and available optimization heuristics. 

The first basic low-level platform-dependent tool~(~\textbf{\emph{ccc-time}}) 
is intended to execute applications on a given architecture, collect various row profile
information such as function level profiling and monitor hardware performance counters
using popular PAPI library~\cite{papi}, PAPIEx~\cite{papiex} and OProfile~\cite{oprofile}
tools. This tool is very small and portable tested on a number of platforms
ranging from supercomputers and high-performance servers based on Intel, AMD, Cell and
NVidia processors and accelerators to embedded systems from ARC, STMicroelectronics and ARM.

Other two major platform-independent components of CCC are
\textbf{\emph{ccc-comp}} and \textbf{\emph{ccc-run}} that provide all the
necessary functionality to compile application with different optimizations,
execute a binary or byte code with multiple datasets, process row profile
statistics from \textbf{\emph{ccc-time}}, validate program output
correctness, etc. Currently, these tools work with specially prepared
programs and datasets such as Collective Benchmark (cBench) described in
section~\ref{sec:tools:cbench} in order to automate the optimization process
and validation of the code correctness. However, the required changes are
minimal and we have already converted all programs from EEMBC, SPEC CPU
95,2000,2006 and cBench to work with CCC framework. We are currently 
extending CCC framework within Google Summer of Code 2009 program~\cite{gsoc2009} 
to enable transparent continuous collective optimization, fine-grain program
transformations and run-time adaptation within GCC without any Makefile
modifications based on a statistical collective optimization concept~\cite{FT2009}.

The command line format of~\emph{ccc-comp} and \emph{ccc-run} is the following:
\begin{itemize}
\item \textbf{ccc-comp} $\langle$descriptive compiler name$\rangle$ $\langle$compiler optimization
flags recorded in COD$\rangle$ $\langle$compiler auxiliary flags not recorded in COD$\rangle$
\item \textbf{ccc-run} $\langle$dataset number$\rangle$ $\langle$1 if baseline reference run (optional)$\rangle$
\end{itemize}

Normally, we start iterative compilation and optimization space exploration
with the baseline reference compilation and execution such
as~\emph{\textbf{ccc-comp}  milepostgcc44  -O3} and~\emph{ccc-run  1  1}
where~\emph{milepostgcc44} is the sample descriptive name of machine learning
enabled GCC with ICI v2.0  and feature extractor v2.0 
registered during CCC configuration of available compilers, 
~\emph{-O3} is the best optimization level of GCC to be improved. The first parameter
of~\emph{ccc-run} is the dataset number (for specially prepared benchmarks
with multiple datasets) and the second parameter indicates that it is the
reference run when the program output will be recorded to help validate
correctness of optimizations during iterative compilation.

\begin{figure}[htbp]

\scriptsize

\#Record compiler passes (through ICI) \\
export CCC\_ICI\_PASSES\_RECORD=1 \\
\\
\#Substitute original GCC pass manager and allow optimization pass selection and reordering (through ICI) \\
export CCC\_ICI\_PASSES\_USE=1 \\
\\
\#Extract program static features when using MILEPOST GCC \\
export CCC\_ICI\_FEATURES\_STATIC\_EXTRACT=1 \\
\#Specify after which optimization pass to extract static features \\
export ICI\_PROG\_FEAT\_PASS=fre \\
\\
\#Profile application using hardware counters and PAPI library \\
export CCC\_HC\_PAPI\_USE=PAPI\_TOT\_INS,PAPI\_FP\_INS,PAPI\_BR\_INS,PAPI\_L1\_DCM,PAPI\_L2\_DCM,PAPI\_TLB\_DM,PAPI\_L1\_LDM \\
\\
\#Profile application using gprof \\
export CCC\_GPROF=1 \\
\\
\#Profile application using oprof \\
export CCC\_OPROF=1 \\
export CCC\_OPROF\_PARAM="--event=CPU\_CLK\_UNHALTED:6000" \\
\\ 
\#Repeat program execution a number of times with the same dataset to detect and remove the performance measurement noise validate stability of execution time statistically\\
export CCC\_RUNS=3\\
\\
\#Architecture specific optimization flags for design space exploration \\
export CCC\_OPT\_PLATFORM="-mA7 -ffixed-r12 -ffixed-r16 -ffixed-r17 -ffixed-r18 -ffixed-r19 -ffixed-r20 -ffixed-r21 -ffixed-r22 -ffixed-r23 -ffixed-r24 -ffixed-r25" \\
export CCC\_OPT\_PLATFORM="-mtune=itanium2" \\
export CCC\_OPT\_PLATFORM="-march=athlon64" \\
\\
\#In case of multiprocessor and multicore system, select which processor/core to run application on \\
export CCC\_PROCESSOR\_NUM=\\
\\
\#Select runtime environment (VM or simulator)\\
export CCC\_RUN\_RE=llvm25\\
export CCC\_RUN\_RE=ilrun\\
export CCC\_RUN\_RE=unisim\\
export CCC\_RUN\_RE=simplescalar\\
\\
\#Some notes to record in COD together with experimental data\\
export CCC\_NOTES="test optimizations"\\
\\
\#The following variables are currently used in the on-going projects and can change:\\
\#Architecture parameters for design space exploration\\
 export CCC\_ARCH\_CFG="l1\_cache=203; l2\_cache=35;"\\
 export CCC\_ARCH\_SIZE=132\\
\#Static parallelization and fine-grain optimizations\\
 export CCC\_OPT\_FINE="loop\_tiling=10;"\\
 export CCC\_OPT\_PAR\_STATIC="all\_loops=parallelizable;"\\
\#Information about power consumption, energy, dynamic dependencies that should be recorded automatically\\
 export CCC\_RUN\_POWER=\\
 export CCC\_RUN\_ENERGY=\\
 export CCC\_PAR\_DYNAMIC="no deps"\\

 \caption{Some environment variables to control \emph{ccc-comp} and \emph{ccc-run} tools from CCC framework}
 \label{fig:ccc_control}
\end{figure}

\begin{figure}[htbp]

\scriptsize

\textbf{Main compilation information packet (local filename:\emph{\_comp}):}\\
\\
COMPILE\_ID=19293849477085514 \\
PLATFORM\_ID=2111574609159278179 \\
ENVIRONMENT\_ID=2781195477254972989 \\
COMPILER\_ID=129504539516446542 \\
PROGRAM\_ID=1487849553352134 \\
DATE=2009-06-04 \\
TIME=14:06:47 \\
OPT\_FLAGS=-O3 \\
OPT\_FLAGS\_PLATFORM=-msse2 \\
COMPILE\_TIME=69.000000 \\
BIN\_SIZE=48870 \\
OBJ\_MD5CRC=b15359251b3c185dfa180e0e1ad16228 \\
ICI\_FEATURES\_STATIC\_EXTRACT=1 \\
NOTES=baseline compilation \\
\\
\textbf{Information packet with ICI optimization passes (local filename:\emph{\_comp\_passes}):}\\
\\
COMPILE\_ID=19293849477085514 \\
COMPILER\_ID=129504539516446542 \\
FUNCTION\_NAME=corner\_draw \\
PASSES=all\_optimizations,strip\_predict\_hints,addressables,copyrename,cunrolli,ccp,forwprop,cdce,alias,retslot,phiprop,fre,copyprop,mergephi,... \\
\\
COMPILE\_ID=19293849477085514 \\
COMPILER\_ID=129504539516446542 \\
FUNCTION\_NAME=edge\_draw \\
PASSES=all\_optimizations,strip\_predict\_hints,addressables,copyrename,cunrolli,ccp,forwprop,cdce,alias,retslot,phiprop,fre,copyprop,mergephi,... \\
...\\
\textbf{Information packet with program features (local filename:\emph{\_prog\_feat}):}\\
\\
COMPILE\_ID=19293849477085514 \\
FUNCTION\_NAME=corner\_draw \\
PASS=fre \\                                                                                       
STATIC\_FEATURE\_VECTOR= ft1=9, ft2=4, ft3=2, ft4=0, ft5=5, ft6=2, ft7=0, ft8=3, ft9=1, ft10=1, ft11=1, ft12=0, ft13=5, ft14=2, ...\\
\\
COMPILE\_ID=19293849477085514 \\
FUNCTION\_NAME=edge\_draw \\
PASS=fre \\
STATIC\_FEATURE\_VECTOR= ft1=14, ft2=6, ft3=5, ft4=0, ft5=7, ft6=5, ft7=0, ft8=3, ft9=3, ft10=3, ft11=2, ft12=0, ft13=11, ft14=1, ...\\
...\\
\\
\textbf{Execution information packet (local filename:\emph{\_run}):}\\
\\
RUN\_ID=22712323769921139 \\
RUN\_ID\_ASSOCIATE=22712323769921139 \\
COMPILE\_ID=8098633667852535 \\
COMPILER\_ID=331350613878705696 \\
PLATFORM\_ID=2111574609159278179 \\
ENVIRONMENT\_ID=2781195477254972989 \\
PROGRAM\_ID=1487849553352134 \\
DATE=2009-06-04 \\
TIME=14:35:26 \\
RUN\_COMMAND\_LINE=1) ../../automotive\_susan\_data/1.pgm output\_large.corners.pgm -c $\rangle$ ftmp\_out \\
OUTPUT\_CORRECT=1 \\
RUN\_TIME=16.355512 \\
RUN\_TIME1=0.000000 \\
RUN\_TIME\_USER=13.822898 \\
RUN\_TIME\_SYS=2.532614 \\
RUN\_PG=\{susan\_corners=12.27,782,0.0156905371\} \\
NOTES=baseline compilation \\

 \caption{Information packets produced by~\emph{ccc-comp} and \emph{ccc-run} tools from CCC framework that are recorded
locally or sent to COD}
 \label{fig:ccc_packets}
\end{figure}

We continue exploration of optimizations invoking~\emph{ccc-comp} and~\emph{ccc-run} tools multiple times with different combinations
of optimizations controlled either through command-line flags or multiple environment variables shown in Figure~\ref{fig:ccc_control}.
At each iterative step these tools compare program output with the output of the baseline reference execution to validate
code correctness (though it is clearly not enough and we would like to provide more formal validation plugins) and prepare
several text files (information packets) with compilation and execution information that are recorded locally and can  
be sent to COD using~\emph{ccc-db-*} tools as shown in Figure~\ref{fig:ccc_packets}. 

Iterative optimization space exploration is performed using high-level plugins that invoke~\emph{ccc-comp} and~\emph{ccc-run}
with different optimization parameters. We produced a few plugins written in C and PHP that implement the following
several search algorithms and some machine learning techniques to predict good optimizations:

\begin{itemize}
\item \textbf{ccc-run-glob-flags-rnd-uniform} - generates uniform random combinations of global optimization
flags (each flag has 50\% probability to be selected for a generated combination of optimizations)
\item \textbf{ccc-run-glob-flags-rnd-fixed} - generates a random combination of global optimizations of
a fixed length
\item \textbf{ccc-run-glob-flags-one-by-one} - evaluate all available global optimizations one by one
\item \textbf{ccc-run-glob-flags-one-off-rnd} - select all optimizations at first step and then remove them one by one (similar
to~\cite{PE2006} one of the modes of the \emph{PathOpt} tool from PathScale compiler suite~\cite{pathscale})
\item \textbf{milepost-gcc} - a wrapper around MILEPOST GCC to automatically extract program features and query cTuning web-service to predict good optimization to improve execution time 
and code size substituting default optimization levels (described more in Section~\ref{sec:scenarios}).
\end{itemize}

When distributing optimization space exploration among multiple users
or on clusters and GRID-like architectures, each user may specify a different random 
seed number to explore different parts of optimization spaces on different
machines. Best performing optimization cases from all users will later be
filtered and joined in COD for further analysis and reuse by the whole community.
For example, we can train machine learning models and predict good optimizations
for a given program on a given architecture using collective optimization data from COD
as shown in Section~\ref{sec:scenarios}.

During iterative compilation we are interested to filter a large amount of obtained data 
and find only those optimization cases that improve execution time, code size, power
consumption and other metrics depending on the user optimization scenarios
or detect some performance anomalies and bugs for further analysis.
Hence, we have developed several platform independent plugins (written in PHP) that analyze data in the local
database, find such optimization cases (for example,~\emph{get-all-best-flags-time} finds
optimization cases that improve execution time and~\emph{get-all-best-flags-time-size-pareto}
find cases that improve both execution time and code size using Pareto-like distribution)
and record them in COD. Continuously updated optimization cases can be viewed  
at~\cite{ctuning_repository}.

When using random iterative search we may obtain complex combinations of optimizations 
without clear indication which particular code transformation improves the code. Therefore,
we can also use additional pruning of each optimization case and remove those optimizations
one by one from a given combination that do not influence performance, code size or other
characteristics using~\emph{ccc-run-glob-flags-one-off-rnd} tool. It helps to improve correlation
between program features and transformations, and eventually improve optimization predictions 
based on machine learning techniques such as decision tree algorithms, for example.

Since we also want to explore dynamic optimizations and architecture designs systematically,
we are gradually extending CCC to support various VM systems such as MONO
and LLVM to evaluate JIT split compilation (finding a balance between static
and dynamic optimization using statistical techniques and machine learning)
and support multiple architecture simulators such as UNISIM, SimpleScalar and others
to enable architecture design space exploration and automate architecture and code
co-optimization. More information about current CCC developments is available
at our collaborative website~\cite{ctuning_ccc}. Some practical CCC usage examples
are presented in Section~\ref{sec:scenarios}.

%%%%%%%%%%%%%%%%%%%%%%%%%%%%%%%%%%%%%%%%%%%%%%%%%%%%%%%%%%%%%%%%%%%%%%%%%%%%%%%%%%%%%%
\subsection{Interactive Compilation Interface}
\label{sec:tools:ici}

In 1999-2002, we started developing memory hierarchy analysis and optimization
tools for real large high-performance applications within MHAOTEU project~\cite{ATAP2000} 
to build the first realistic adaptive compiler (follow up of the OCEANS project~\cite{europar97}).
Within MHAOTEU project, we attempted to generalize iterative feedback-directed compilation and 
optimization space exploration to adapt any program to any architecture empirically and
automatically improving execution time over the best default optimization heuristic
of the state-of-the-art compilers. We decided to focus on a few well-known
loop and data transformations such as array padding, reordering and
prefetching, loop tiling (blocking), interchange, fusion/fission,
unrolling and vectorization as well as some polyhedral transformations. Unlike~\cite{CSS99,CST02} 
where only optimization orders have been evaluated on some small kernels using architecture simulator, 
we decided to use large SPEC95 and SPEC2000 floating point benchmarks together with a few real 
applications from MHAOTEU partners as well as several modern at that time architectures 
to evaluate our approach in practice.

Unfortunately, at that time we could not find any production compiler with
fine-grain control of optimization selection while several available source-to-source
transformation tools including SUIF~\cite{suif_compiler} were not 
stable enough to parse all SPEC codes or enable systematic exploration of complex 
combinations of optimizations. Hence, we decided to develop our own source-to-source
compiler and iterative compilation infrastructure~\cite{eos} using Octave
C/C++/Fortran77 front-end and MARS parallelizing compiler based on polyhedral
transformation framework produced at Manchester and Edinburgh Universities~\cite{O1998}. 
However, it was a very tedious and time-consuming task allowing us to evaluate evaluate iterative
compilation using only loop tiling, unrolling and array padding by the end of the project. 
Nevertheless, we got encouraging execution time improvements for SPEC95 and several real
large codes from our industrial partners across several RISC and CISC
architectures~\cite{FOK02,Fur2004}. We also managed to develop a prototype of
quick upper-bound performance evaluation tool as a stopping criteria for
iterative compilation~\cite{Fur2004,FOTP04}.

MHAOTEU project helped us to highlight multiple problems when building adaptive compilers
and indicated further research directions. For example, state-of-the-art static compilers
often fail to produce good quality code (to achieve better code size, execution
time, etc) due to hardwired ad-hoc optimization heuristics (cost models) 
on rapidly evolving hardware, large irregular optimization spaces, fixed order and
complex interaction between optimizations inside compiler or between compiler
and source-to-source or binary transformation tools, time-consuming
retuning of default optimization heuristic for all available architectures
when adding new transformations, inability to retarget compiler
easily for new architectures particularly during architecture design space exploration,
inability to produce mixed-ISA code easily, inability to reuse optimization knowledge 
among different programs and architectures, lack of run-time information,
inability to parallelize code effectively and automatically, 
lack of run-time adaptation mechanisms for statically compiled programs to be
able to react to varying program and system behavior as well
as multiple datasets (program inputs) with low overhead. 
To overcome these problems, we decided to start a long term project to
change outdated compilation and optimization technology radically 
and build a novel realistic adaptive optimization infrastructure that allows 
rigorous systematic evaluation of empirical iterative compilation, run-time adaptation, 
collective optimization, architecture design space exploration and machine learning.

\begin{figure}[htb]
 \centering
 \includegraphics[width=6in]{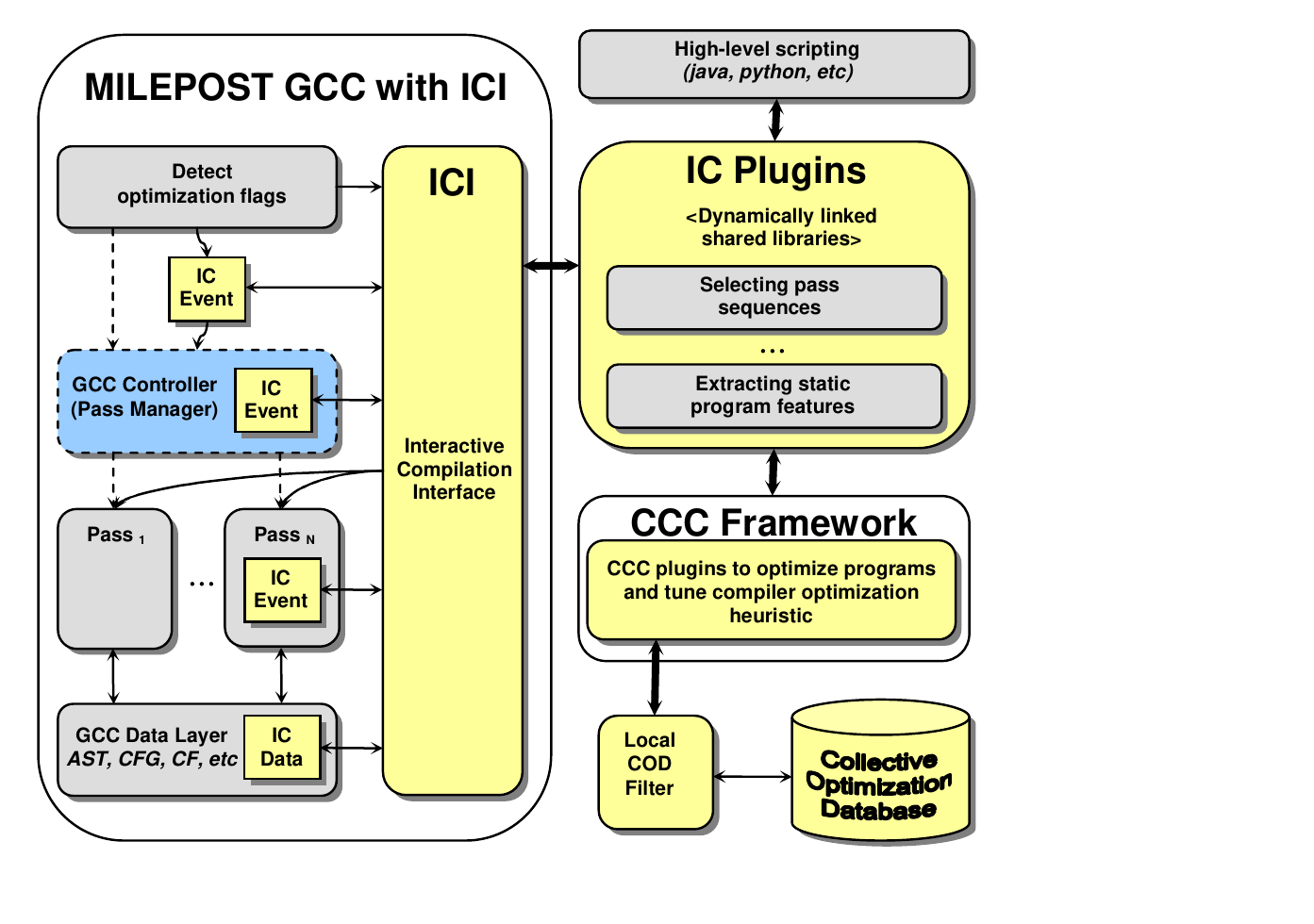}
 \caption{Interactive Compilation Interface: a high-level event-driven plugin framework to open up 
  compilers, extend them and control their internal decisions using dynamically loaded user plugins. 
  This is the first step to enable future modular self-tuning adaptive compilers.}
 \label{fig:ici}
\end{figure}

First, we had to decide which program transformation tool to use to enable systematic
performance evaluation of our research optimization techniques, i.e. continue developing our 
own source-to-source interactive compiler which is too time consuming or find some other solution. 
At the same time, we noticed that some available open-source production compilers such
as Open64 started featuring many aggressive loop and array transformations that
we planned to implement in our transformation tool. Considering that Open64 was a stable compiler
supporting two architectures, C/Fortran languages and could process most of the SPEC codes,
we decided to use it for our experiments. We provided a simple interface to open it up and
enable external selection of internal loop/array optimizations and their parameters through an event-driven
plugin system that we called Interactive Compilation Interface (ICI). We combined
it with the Framework for Continuous Optimizations (FCO) and developed several 
search plugins (random, exhaustive, leave one out, hill climbing) to enable continuous and 
systematic optimization space exploration using ICI-enabled compiler~\cite{fco}. 
Since we released ICI for Open64/PathScale compiler in 2005, it has proved to be a simple and
efficient way to transform production compilers into iterative research tools and
was used in several  research projects to fine-tune programs for a given architecture and
a dataset~\cite{FCOP2005,LCWP2009}. However, when we tried to extend it to combine
reordering of optimizations with fine grain optimizations, we found it too
time consuming to modify the rigid optimization manager in Open64. 

At the same time, we noticed that there was a considerable community effort to modularize GCC,
add a new optimization pass manager with some basic information about dependencies
between passes and provide many aggressive optimizations including polyhedral transformations.
Considering that it could open up many interesting research opportunities
and taking into account that GCC is a unique stable open-source 
compiler that supports dozens of various architectures, multiple languages, 
can compile the whole Linux and has a very large community that is important for collective
optimization~\cite{FT2009}, we decided to use this compiler for our further research. 
We developed a new ICI to "hijack" GCC and control its internal decisions
through event-driven mechanisms and dynamically loaded plugins.
The concept of the new ICI and interactive compilers has been described in~\cite{FC2007} 
and extended during the MILEPOST project~\cite{FMTP2008}. 
Since then, we have moved all the developments to the community-driven website~\cite{ctuning_ici}
and continued extending it based on user feedback and our research requirements.

Current ICI is an event-driven plugin framework with a high-level compiler-independent and low-level compiler-dependent API
to transform production compilers into collaborative open modular interactive toolsets as shown in figure~\ref{fig:ici}. 
ICI acts as a "middleware" interface between the compiler and the user plugins
and enables external program analysis and instrumentation, fine-grain program optimizations without revealing
all the internals of a compiler. This makes such ICI-enabled compilers more researchers/developers 
friendly allowing simple prototyping of new ideas without a deep knowledge of a compiler itself,
without a need to recompile a compiler itself and avoiding building new compilation and optimization tools 
from scratch. Using ICI can also help to avoid time consuming revolutionary approaches to create 
a new "clean", modular and fast compiler infrastructure from scratch while gradually
transforming current rigid compilers into a modular adaptive compiler infrastructure.
Finally, we believe that using production compilers with ICI in research can help to move
successful ideas back to the compiler much faster for the benefit of all users
and boost innovation and research.

\begin{figure}[htb]
 \scriptsize

 \centering
  {\footnotesize
  \begin{tabular}{|l|l|p{220pt}|}
    \hline
    \textbf{Feature name:} & \textbf{Type of contents:} & \textbf{Purpose} \\
    \hline
      compiler\_flags          & array of strings (char **)  & Names of all
      known command-line options (flags) of a compiler. Individual option
      names are stored without the leading dash. \\

      compiler\_params         & array of strings (char **)  & Names of all known compiler parameters. \\

      function\_name           & string (char)               & Name of the function currently being compiled.  \\

      function\_decl\_filename & string (char)               & Name of the
      file in which the function currently being compiled was declared.
      Returns the filename corresponding to the most recent declaration.  \\

      function\_decl\_line     & integer (int)               & Line number
      at which the function currently being compiled was declared. In
      conjuction with feature "function\_decl\_filename" gives the location
      of the most recent declaration of the current function. \\

      function\_filename       & string (char)               & Name of the
      file in which the function currently being compiled was defined.  \\

      function\_start\_line    & integer (int)               & Line number
      at which the definition of the current function effectively starts.
      Corresponds to the first line of the body of the current function.  \\

      function\_end\_line      & integer (int)               & Line number
      at which the definition of the current function effectively ends.
      Corresponds to the last line of the body of the current function.  \\

      first\_pass              & string (char)               &
      Human-readable name of the first pass of a compiler. Accessing this
      feature has the side effect of setting that specific pass as the
      "current pass" of ICI.  \\

      next\_pass               & string (char)               &
      Human-readable name of the next pass to be executed after the "current
      pass" of ICI. Accessing this feature has the side effect of advancing
      the "current pass" of ICI to its immediate successor in the currently
      defined pass chain.  \\
    \hline
  \end{tabular}
 }

 \caption{List of some popular features available in ICI version 2.x}
 \label{fig:ici:features}
\end{figure}

We used GCC with ICI in the MILEPOST project~\cite{milepost} to
develop the first machine learning enabled research compiler
and enable automatic and systematic optimization space exploration and predict good
combinations of optimizations (global flags or passes on a function level)
based on static program features and predictive modeling. Together with
colleagues from IBM we could easily add feature extractor pass to GCC
and had an ability to call it after any arbitrary optimization pass using
simple dynamic plugin. Such machine learning enabled self-tuning compiler
called MILEPOST GCC (GCC with ICI and static feature extractor) has been released 
and used by IBM and ARC to improve and speed up the optimization process
for their realistic applications. This usage scenario is described more in Section~\ref{sec:scenarios})
and in~\cite{FMTP2008}.

Current ICI is organized around four main concepts to abstract the compilation process:
\begin{itemize}
\item \textbf{Plugins} are dynamically loaded user modules that "hijack" a
compiler to control compilation decisions and have an access to some or all
of its internal functions and data. Currently, the plugin programming
interface consists of three kinds of functions:~\texttt{initialization
function} that is in charge of starting the plugin, checking compiler and
plugin compatibility, and registering event handlers;~\texttt{termination
function} that is in charge of cleaning up the plugin data structures
and closing files, etc;~\texttt{event handler (callback) functions} that control
a compiler.
\item \textbf{Events} are triggered whenever a compiler reaches some defined
point during execution. In such case, ICI invokes a user-definable callback
function (event handler) referenced simply by a string name.
\item \textbf{Features} are abstractions of selected properties of a current
compiler state and of a compiled program. The brief list of some available
features is shown in Figure~\ref{fig:ici:features} ranging from an array of
optimization passes to simple string name of a compiled function.
\item \textbf{Parameters} are abstractions of compiler variables to 
decouple plugins from the actual implementation of compiler internals.
They are identified simply by a string name and used to get and/or set some
values in the compiler such as force inlining of some function or change loop
blocking or unrolling factors, for example.
\end{itemize}

Detailed documentation is available at the ICI collaborative website~\cite{ctuning_ici_docs}.

Since 2007, we have been participating in multiple discussions with other
colleagues developing their own GCC plugin frameworks such
as~\cite{Sta2007,GM2008,CDZ2007}, GCC community and steering committee to
add a generic plugin framework in GCC. Finally, a plugin framework will be
included in mainline GCC 4.5~\cite{gcc_plugin_framework}. This plugin
framework is very similar to ICI but more low-level. For example, the set of
plugin events is hardwired inside a compiler, plugin callbacks have a fixed,
pass-by-value argument set and the pass management is very basic. However,
it already provides a multi-plugin support with command-line arguments and
callback chains, i.e. lists of callbacks invoked upon a single occurrence
of a plugin event, and is a good step towards interactive adaptive compilers. 
Hence, we are synchronizing ICI with the plugin branch~\cite{gcc_ici_sync} to
provide more high-level API including:

\begin{itemize}
\item dynamic registration and unregistration of plugin events
\item dynamic registration/definition/unregistration of event callback arguments
\item arbitrary number of pass-by-name event callback arguments
\item ability to substitute complete pass managers (chains of passes) 
\item high-level access to a compiler state (values of flags and parameters, name and selected properties 
of the current function, name of a current and next pass) with some modification possibilities (compiler 
parameters, next pass).
\end{itemize}

Comparison of ICI and some other available plugin framework for GCC is available
at~\cite{gcc_plugin_comparisons}. ICI plugin repository with several test,
pass manipulation and machine learning plugins is available at the
collaborative development website~\cite{ctuning_ici}. During Google Summer
of Code'09 program~\cite{gsoc2009} we have extended ICI and plugins to provide
XML representation of the compilation flow, selection and tuning of
fine-grain optimizations/polyhedral GRAPHITE transformations and their
parameters using machine learning, enable code instrumentation, generic
function cloning, run-time adaptation capabilities and collective
optimization technology~\cite{FT2009}. We also ported ICI and MILEPOST
program feature extractor to GCC4NET~\cite{gcc4net} to evaluate
split compilation, i.e. predicting the good balance between optimizations that
should be performed at a compile time and the ones that should be performed at
run-time when executing code on multiple architectures and with multiple
datasets based on statistical analysis and machine learning.

We hope that ICI-like plugin framework will become standard for compilers in the future,
will help prototype research ideas quickly, will simplify, modularize and automate
compiler design, will allow users to write their own optimization plugins, 
will enable automatic tuning of optimization heuristic and retargetability
for different architectures and will eventually enable smart self-tuning adaptive 
computing systems for the emerging heterogeneous (and reconfigurable) multi-core architectures.
More information about ICI and current collaborative extension projects
is available at~\cite{ctuning_ici}.

%%%%%%%%%%%%%%%%%%%%%%%%%%%%%%%%%%%%%%%%%%%%%%%%%%%%%%%%%%%%%%%%%%%%%%%%%%%%%%%%%%%%%%
\subsection{Collective Benchmark}
\label{sec:tools:cbench}

Automatic iterative feedback-directed compilation is now becoming a standard technique to optimize
programs, evaluate architecture designs and tune compiler optimization heuristics. However, it is
often performed with one or several datasets (test, train and ref in SPEC benchmarks for example)
with an implicit assumption that the best configuration found for a given program using one or several datasets
will work well with other datasets for that program. 

We already know well that different optimizations are needed for different datasets when optimizing small
kernels and libraries~\cite{atlas,fftw,spiral,europar97}. For example, different tiling and unrolling factors
are required for matrix multiply to better utilize memory hierarchy/ILP and improve execution time depending 
on the matrix size. However, when evaluating iterative feedback-directed fine-grain optimizations 
(loop tiling, unrolling and array padding) for large applications even with one dataset~\cite{FOK02,Fur2004} 
we confirmed that the effect of such optimizations on a large code can be very different then on kernels 
and normally much smaller often due to inter-loop and inter-procedural memory locality, complex data dependencies 
and low memory bandwidth among others.

In order to enable systematic exploration of iterative compilation
techniques and realistic performance evaluation for programs with multiple datasets 
we need benchmarks that have a large number of inputs together with tools that support global 
inter-procedural and coarse-grain optimizations (and parallelization)
based on combination of traditional fine-grain optimizations and polyhedral
transformations. 

Unfortunately, most of the available open-source and commercial benchmarks
has only a few datasets available. Hence, in 2006, we decided to assemble a collection of data sets
for a free, commercially representative MiBench~\cite{miBench} benchmark suite.
Originally, we assembled 20 inputs per program, for 26 MiBench programs (520 data sets in total)
in the dataset suite that we called MiDataSets~\cite{FCOP2007}. We started from the top-down
approach evaluating first global optimizations (using compiler flags)~\cite{FCOP2007} 
and gradually adding support to evaluate individual transformations including polyhedral 
GRAPHITE optimizations in GCC using Interactive Compilation Interface withing
GSOC'09 program~\cite{gsoc2009}. 

We released MiDataSets in 2007 and since then it has been used in multiple
research projects. Therefore, we decided to extend it, add more programs and
kernels, update all current MiBench programs to support ANSI C and improve
portability across multiple architectures, and create a dataset repository.
Therefore, we developed a new research benchmark called Collective Benchmark
(cBench) with an open repository~\cite{ctuning_cbench} to keep various
open-source programs with multiple inputs assembled by the community. 

Naturally, the span of execution times across all programs and all datasets
can be very large which complicates systematic iterative compilation evaluation.
For example, when the execution time is too low, it may be difficult to evaluate the impact
of optimizations due to measurement noise. In this cases, we add a loop wrapper around 
a main function moving most of the IO and initialization routines out of it to be able 
to control the length of the program execution. A user can change the upper bound of the loop 
wrapper through environment variable or a special text file. We provide the default setting that
makes a program run about 10 seconds on AMD Athlon64 3700+. However, if execution time 
is too high and slows down systematic iterative compilation particularly when 
using architecture simulators, we are trying to detect those program variables 
that can control the length of the program execution and allow users to modify
them externally. Of course, in such cases, the program behavior may change due to 
locality issues among others and may require different optimizations which is a subject 
of further research.

Each program from cBench currently has several Makefiles for different compilers
including GCC, GCC4CLI, LLVM, Intel, Open64, PathScale (Makefile.gcc, Makefile.gcc4cli,
Makefile.llvm, Makefile.intel, Makefile.open64, Makefile.pathscale respectively).
Two basic scripts are also provided to compile and run a program:

\begin{itemize}
\item \textbf{\_\_compile} $\langle$Makefile compiler extension$\rangle$ $\langle$Optimization flags$\rangle$
\item \textbf{\_\_run} $\langle$dataset number$\rangle$ ($\langle$loop wrapper bound - using default number if omitted$\rangle$)
\end{itemize}

Datasets are described in file~\textbf{\_ccc\_info\_datasets} that has the following format:\\
\texttt{
\\
$\langle$Total number of available datasets$\rangle$ \\
====\\
$\langle$Dataset number$\rangle$\\
$\langle$Command line when invoking executable for this dataset$\rangle$\\
$\langle$Loop wrapper bound$\rangle$\\
====\\
...
}

Since one of the main purposes of cBench is enabling rigorous systematic evaluation 
of empirical program optimizations, we included several scripts and files 
for each cBench program to work with CCC framework and record all experimental 
data in COD entirely automatically. These scripts and files include:

\begin{itemize}
\item \textbf{\_ccc\_program\_id} - file with unique CCC framework ID to be able to share
optimization cases with the community within COD~\cite{ctuning_repository}.
\item \textbf{\_ccc\_prep} - script that is invoked before program compilation
to prepare directory for execution, i.e. copying some large datasets
or compile libraries, for example. It is used for SPEC2006 for example.
\item \textbf{\_ccc\_post} - script that is invoked after program execution
and can be useful when copying profile statistics from remote servers. For example,
it is used when executing programs remotely on ARC simulation board using SSH.
\item \textbf{\_ccc\_check\_output.clean} - script that removes all output files a program may produce.
\item \textbf{\_ccc\_check\_output.copy} - script that saves all output files after a reference run.
\item \textbf{\_ccc\_check\_output.diff} - script that compares all output
files after execution with the saved outputs from the reference run to have
a simple check that a combination of optimizations have been correct. Of course,
this method does not prove correctness and we plan to add more formal methods, 
but it can quickly identify bugs and remove illegal combinations of optimizations.
\end{itemize}

We believe that this community-assembled benchmark with multiple dataset
opens up many research opportunities for realistic code and architecture
optimization, improves the quality and reproducibility of the systematic
empirical research and can enable more realistic benchmarking of computing systems.
For example, we believe that using just one performance metric produced by
current benchmarks with some ad-hoc combinations of optimizations and a few
datasets may not be enough to characterize the overall behavior of the system
since using iterative optimization can result in a much better code.
Our approach is to enable continuous monitoring, optimization and characterization
of computing systems (programs, datasets, architectures, compilers) to be able to 
provide a more realistic performance upper bound and comparison after iterative compilation.

We also plan to extend cBench by extracting most time-consuming kernels
(functions or loops) from programs with the snapshots of their 
inputs during multiple program phases (such kernels with encapsulated inputs
are called codelets). We will randomly modify and combine them together to 
produce large training sets from kernels with various features
in order to answer a research question whether it is possible to predict good optimizations
for large programs based on program decomposition and kernel optimizations.
Finally, we plan to add parallel programs and kernels (OpenCL, OpenMP, CUDA, MPI, etc)
with multiple datasets to extend research on adaptive parallelization
and scheduling for programs with multiple datasets for heterogeneous 
multi-core architecture~\cite{JGVP2009}. 

It is possible to download cBench and participate in collaborative development at~\cite{ctuning_cbench}.

%%%%%%%%%%%%%%%%%%%%%%%%%%%%%%%%%%%%%%%%%%%%%%%%%%%%%%%%%%%%%%%%%%%%%%%%%%%%%%%%%%%%%%
\subsection{Universal run-time Adaptation Framework for Statically-Compiled Programs}
\label{sec:tools:unidapt}

\begin{figure}[htb]
 \centering
 \includegraphics[width=6in]{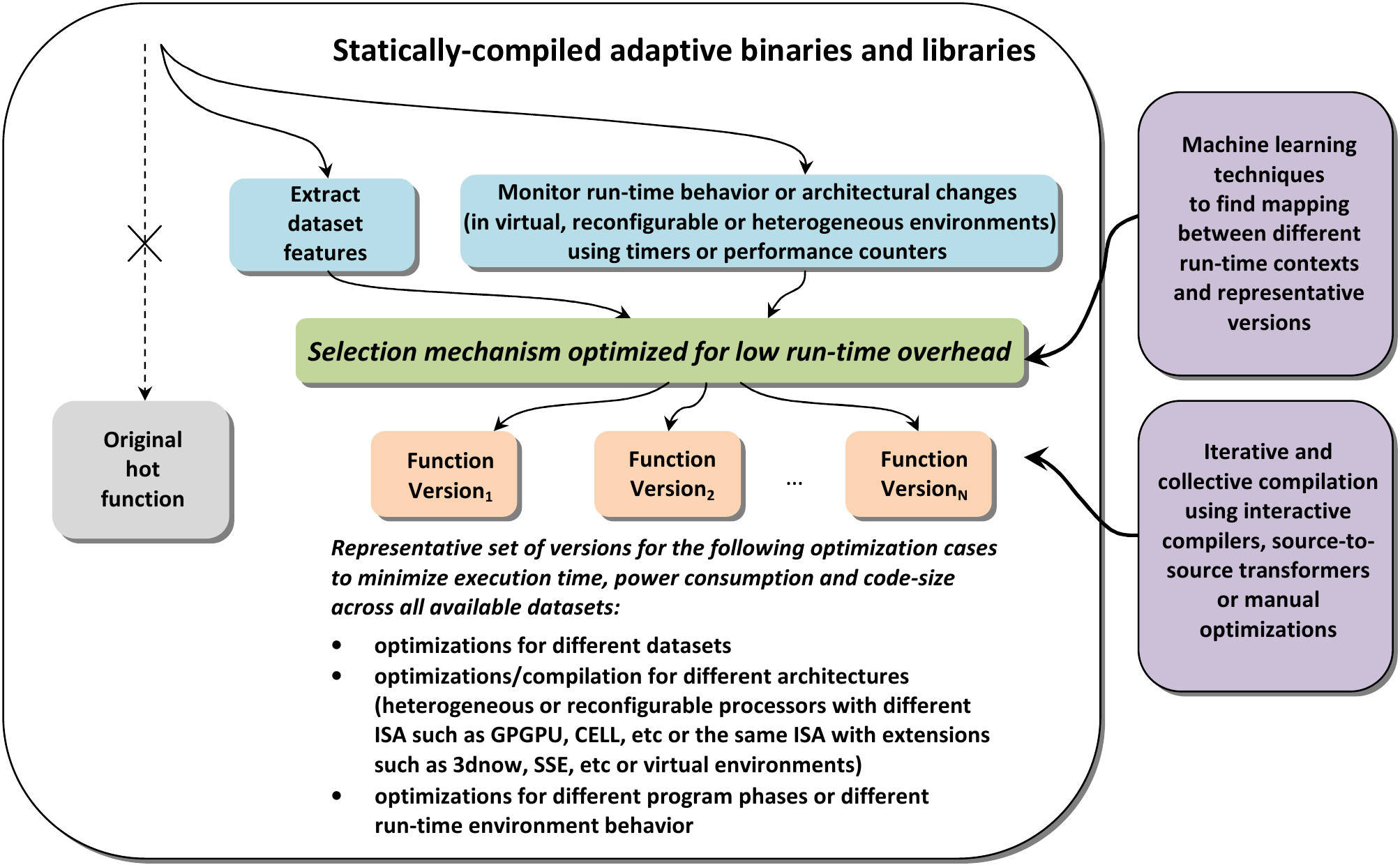}
 \caption{Universal run-time adaptation framework (UNIDAPT) based on static multiversioning and dynamic selection routines.
A representative set of multiple function versions~\cite{LCWP2009} optimized or compiled for different run-time 
optimization scenarios is used in such adaptive binaries and libraries. An optimized decision tree
or rule induction techniques should be able to map clones to different run-time contexts/scenarios
during program or library execution.}
 \label{fig:unidapt}
\end{figure}

As already mentioned in previous sections, iterative compilation can help optimize programs
for any given architecture automatically, but often performed with only one or several datasets. 
Since it is known that optimal selection of optimizations may depend on program inputs~\cite{atlas,fftw,spiral,FCOP2007},
just-in-time or hybrid static/dynamic optimization approaches have been introduced
to select appropriate optimizations at run-time depending on the context and optimization scenario.
However, it is not always possible to use complex recompilation framework particularly
in case of resource-limited embedded systems. Moreover, most of the available systems
are often limited to only a few optimizations and do not have mechanisms to select a
representative set of optimization variants~\cite{BDHP1987,DR1997,VE2001,LAHP2006}. 

In~\cite{FCOP2005} we presented a framework (UNIDAPT) that enabled run-time
optimization and adaptation for statically compiled binaries and libraries
based on static function multiversioning, iterative compilation and
low-overhead hardware counters monitoring routines. During compilation,
several clones of hot functions are created and a set of optimizations is
applied to the clones that may improve execution time across a number of
inputs or architecture configurations. During execution, UNIDAPT framework
monitors original program and system behavior and occasionally invokes
clones to build a table that associates program behavior based on hardware
counters (dynamic program features) with the best performing clones, i.e.
optimizations. This table is later used to predict and select good clones as
a reaction to a given program behavior. It is continuously updated during
executions with multiple datasets and on multiple architectures thus
enabling simple and effective run-time adaptation even for statically-compiled 
applications. During consequent compilations the worst performing clones 
can be eliminated and new added to enable continuous and transparent 
optimization space exploration and adaptation. Similar framework has been also recently
presented in~\cite{MH2009}.

Our approach is driven by simplicity and practicality. We show that with
UNIDAPT framework it is possible to select complex optimizations at run-time
without resorting to sophisticated dynamic compilation frameworks. 
Since 2004, we have extended this approach in multiple research projects. 
We used it to speed up iterative compilation by several orders of magnitude~\cite{FCOP2005} using
low-overhead program phase detection at run-time; evaluate run-time
optimizations for irregular codes~\cite{FMPP2007}; build self-tuning
multi-versioning libraries automatically using representative sets of
optimizations found off-line and providing fast run-time selection routines
based on dataset features and standard machine learning
techniques such as decision tree classifiers~\cite{LCWP2009}; enable transparent statistical continuous
collective optimization of computing systems~\cite{FT2009}. We also started
investigating predictive run-time code scheduling for heterogeneous
multi-core architecture where function clones are targeted for different ISA
together with explicit data management~\cite{JGVP2009}. 

Since 2006, we are gradually implementing UNIDAPT framework presented 
in Figure~\ref{fig:unidapt} in GCC. During Google Summer of
Code'09 program~\cite{gsoc2009} we are extending GCC to generate
multiple function clones on the fly using ICI, apply combinations of
fine-grain optimizations to the generated clones, provide program
instrumentation to call program behavior monitoring and adaptation routines
(and explicit memory transfer routines in case of code generation for
heterogeneous GPGPU-like architectures~\cite{JGVP2009}), provide transparent
linking with external libraries (to enable monitoring of hardware counters
or machine learning libraries to associate program behavior with optimized
clones, for example) and add decision tree statements to select appropriate
clones at run-time according to dynamic program features.

UNIDAPT framework combined with cTuning infrastructure opens up many research opportunities.
We are extending it to improve dataset, program behavior and architecture characterization
to better predict optimizations; provide run-time architecture reconfiguration to improve both execution 
time and power consumption using program phases based on~\cite{FCOP2005};
enable split-compilation, i.e. finding a balance between static and dynamic optimizations
using predictive modeling; improve dynamic parallelization, data partitioning, caching, 
and scheduling for heterogeneous multi-core architectures; 
enable migration of an optimized code in virtual environments when architecture may change at run-time;
provide fault-tolerance mechanisms by adding clones compiled with soft-error correction mechanisms,for example.

Finally, we started combining cTuning/MILEPOST technology, UNIDAPT framework 
and a Hybrid Multi-core Parallel Programming Environment (HMPP)~\cite{hmpp}
to enable adaptive practical profile-driven optimization and parallelization for the current and future 
hybrid heterogeneous CPU/GPU-like architectures based on dynamic collective optimization, dynamic
data partitioning and predictive scheduling, empirical iterative compilation, statistical analysis, 
machine learning and decision trees together with program and dataset features
~\cite{FT2009,LCWP2009,JGVP2009,TWFP2009,FMTP2008,LFF2007,FCOP2005}. 

More information about collaborative UNIDAPT R\&D is available at~\cite{ctuning_unidapt}.

%%%%%%%%%%%%%%%%%%%%%%%%%%%%%%%%%%%%%%%%%%%%%%%%%%%%%%%%%%%%%%%%%%%%%%%%%
\section{Usage Scenarios}
\label{sec:scenarios}

\subsection{Manual sharing of optimization cases}

Collective tuning infrastructure provides multiple ways to optimize 
computing systems and opens up many research opportunities.
In this section we will present several common usage scenarios.

The first and the simplest scenario is manual sharing and reuse of optimization cases
using an online web form at~\cite{ctuning_repository}. If a user finds some 
optimization configuration such as a combination of compiler flags, order of optimization passes,
parameters of fine-grain transformations, architecture configuration, etc
that improves some characteristics of a given program such as execution time, code size, power consumption, 
rate of soft errors, etc over default compiler optimization level and default architecture
configuration with a given dataset, such optimization case can be submitted
to Collective Optimization Database to make the community aware of it.

\begin{figure}[htb]
  \centering
  \includegraphics[width=5in]{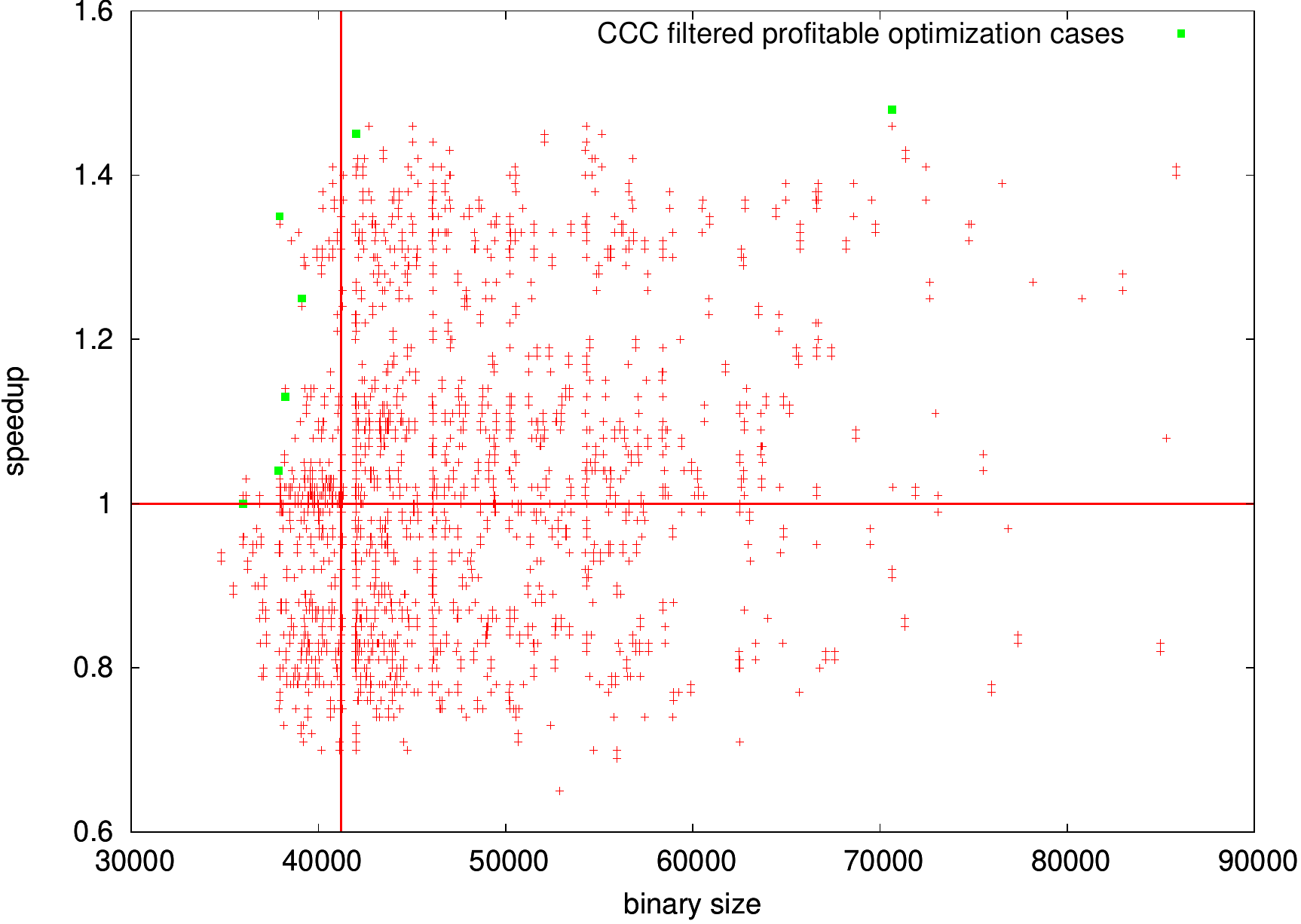}
  \caption{\it Variation of speedups and size of the binary for~\emph{susan\_corners} using GCC 4.2.2 on AMD Athlon64 3700+ 
during automatic iterative feedback-directed compilation performed by CCC framework 
over best available optimization level (-O3) and different profitable optimization cases detected by CCC filter plugins depending on optimization scenarios (similar to Pareto distribution).}
  \label{fig:example_susan_c_amd64}
\end{figure}

If an optimized program and a dataset are well-known from standard benchmarks such 
as SPEC, EEMBC, MiBench/cBench/MiDataSets or from open-source projects, they can be 
simply referenced by their name, benchmark suite and a version. If a program 
is less known in the community, have open sources and can be distributed without any limitations, 
a user may be interested to share it with the community together with multiple datasets 
within our Collective Benchmark (cBench) at~\cite{ctuning_cbench}
to help other users reproduce and verify optimization cases. This can be particularly
important when using COD to reference bugs in compilers, run-time systems, architecture simulators
and other components of computing systems. However, if a program/dataset pair is not an open source
while a user or a company would still like to share optimization cases for it with the community
or get an automatic optimization suggestion based on collective optimization knowledge,
such pair can be characterized using static or dynamic features and program reaction to 
transformations that can capture code and architecture characteristics without revealing the sources 
to be able to compare programs and datasets indirectly~\cite{ABCP06,FMTP2008,CFAP2007,LCWP2009,FT2009}
(on-going work).

When optimizing computing systems, users can browse COD to find optimization cases
for similar architectures, compilers, run-time environments, programs and datasets. 
Users can reproduce and improve optimization cases, provide notes
and rank cases to favor the best performing ones. We plan to automate ranking
of optimization cases eventually based on statistical collective optimization concept~\cite{FT2009}.

Collective tuning infrastructure also helps to improve the quality of
academic and industrial research on code, compiler and architecture design
and optimization by enabling open characterization, unique referencing and
fair comparison of the empirical optimization experiments. It is thus intended to address
one of the drawbacks of academic research when it is often difficult or
impossible to reproduce prior experimental results. We envision that authors
will provide experimental data in COD when submitting research papers or
after publication to allow verification and comparison with available
techniques. Some data can be marked as private and accessible only by
reviewers until the paper is published.

When it is not possible to describe optimization cases using current COD
structure, a user can record information in temporal extension fields using
XML format. If this information is considered important by the community,
the COD structure is continuously extended to keep more information in
permanent fields. Naturally, we use a top-down approach for COD first providing
capabilities to describe global and coarse-grain program optimizations and
then gradually adding fields to describe more fine-grain optimizations. 

\subsection{Automatic and systematic optimization space exploration and benchmarking}

One of the key features of cTuning infrastructure is the possibility to
automate program and architecture optimization exploration using empirical
feedback-directed compilation and to share profitable optimization cases
with the community in COD~\cite{ctuning_repository}. This enables faster 
distributed collective optimization of computing systems and reduces 
release time for new programs, compilers and architectures considerably.

\begin{figure}[htb]
  \centering
  \includegraphics[width=5.6in]{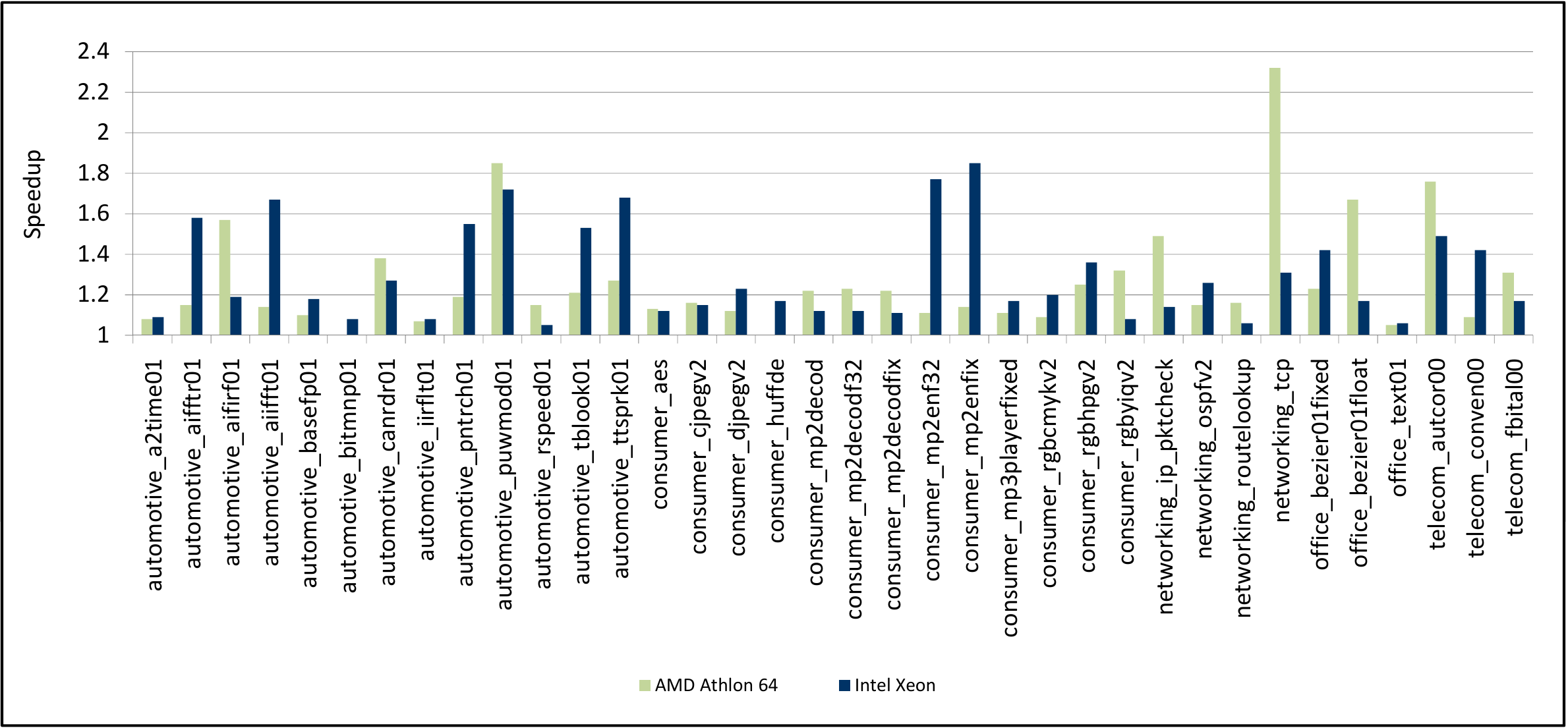} \\
  EEMBC v1.0 and v2.0 \\
  \includegraphics[width=5.6in]{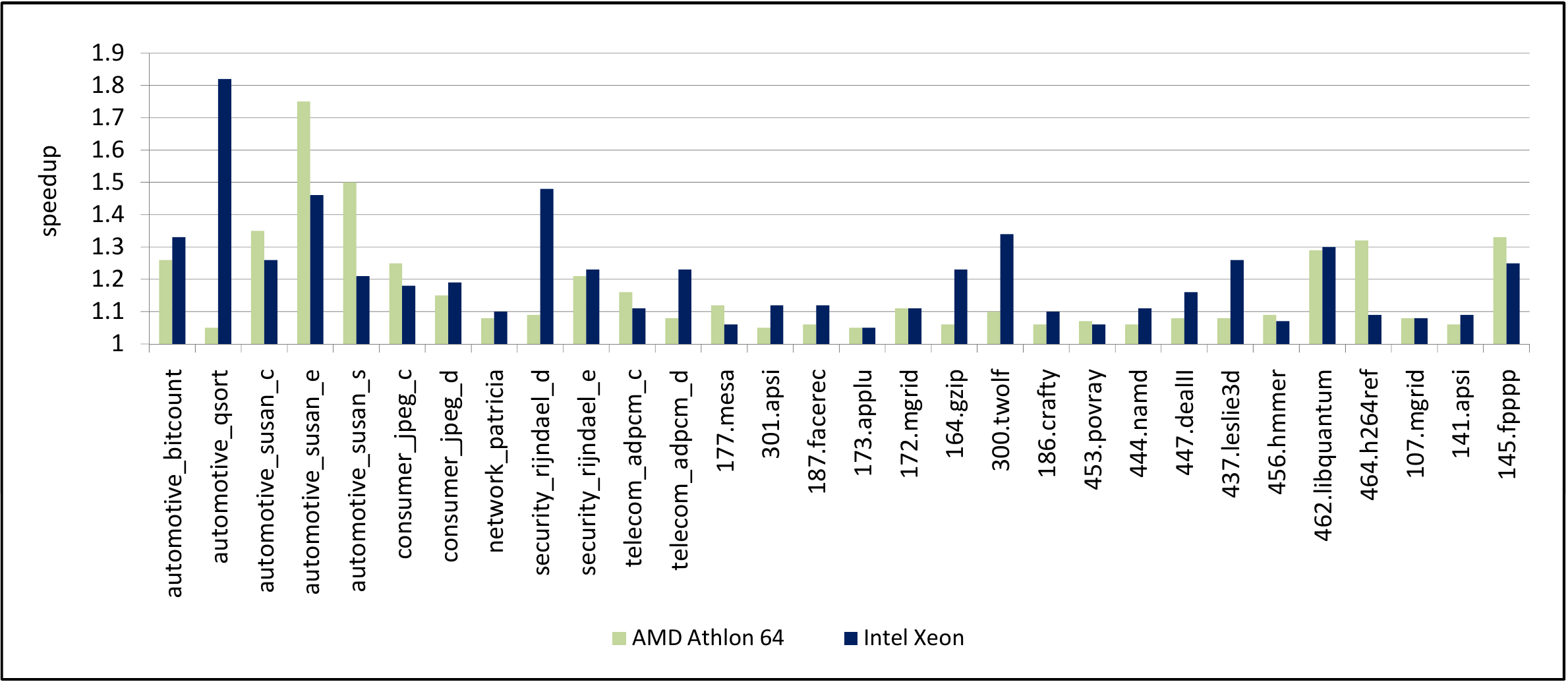} \\
  cBench v1.0, SPEC95,2000,2006 \\
  \caption{\it Evaluation of mature GCC 4.2.2 using iterative compilation with uniform random distribution (500 iterations)
on 2 distinct architectures.}
  \label{fig:all_speedups}
\end{figure}

CCC framework~\cite{ctuning_ccc} is used to automate and distribute
multi-objective code optimization to improve execution time and code size
among other characteristics for a given architecture using multiple search
plugins including exhaustive, random, one off, hill-climbing and other
strategies. Figure~\ref{fig:example_susan_c_amd64} shows distribution of optimization points 
in the 2D space of speedups vs code size of~\emph{susan\_corners} from 
cBench/MiBench~\cite{ctuning_cbench} on AMD Athlon64 3700+ architecture
with GCC 4.2.2 during automatic program optimization using
CCC \texttt{ccc-run-glob-flags-rnd-uniform} plugin after 500 uniform random
combinations of more than 100 global compiler flags (each flag has 50\%
probability to be selected for a given combination of optimizations). 
Naturally, it can be very time consuming and difficult to find good optimization cases
manually in such a non-trivial space and particularly during multi-objective
optimizations. Moreover, the search often depends on optimization
scenario, i.e. it is critical to produce the fastest for high-performance 
servers and supercomputers while it can be more important to find a good balance
between execution time and code size for embedded systems or adaptive libraries.
Hence, we developed several CCC filtering plugins to select optimal
optimization cases for a given program, dataset and architecture based on Pareto-like distributions~\cite{HB2006,HE2008} 
(shown by square dots in Figure~\ref{fig:example_susan_c_amd64}, for example)
before sharing them with the community in COD.

\begin{figure}[htb]
  \centering
  \includegraphics[width=6.5in]{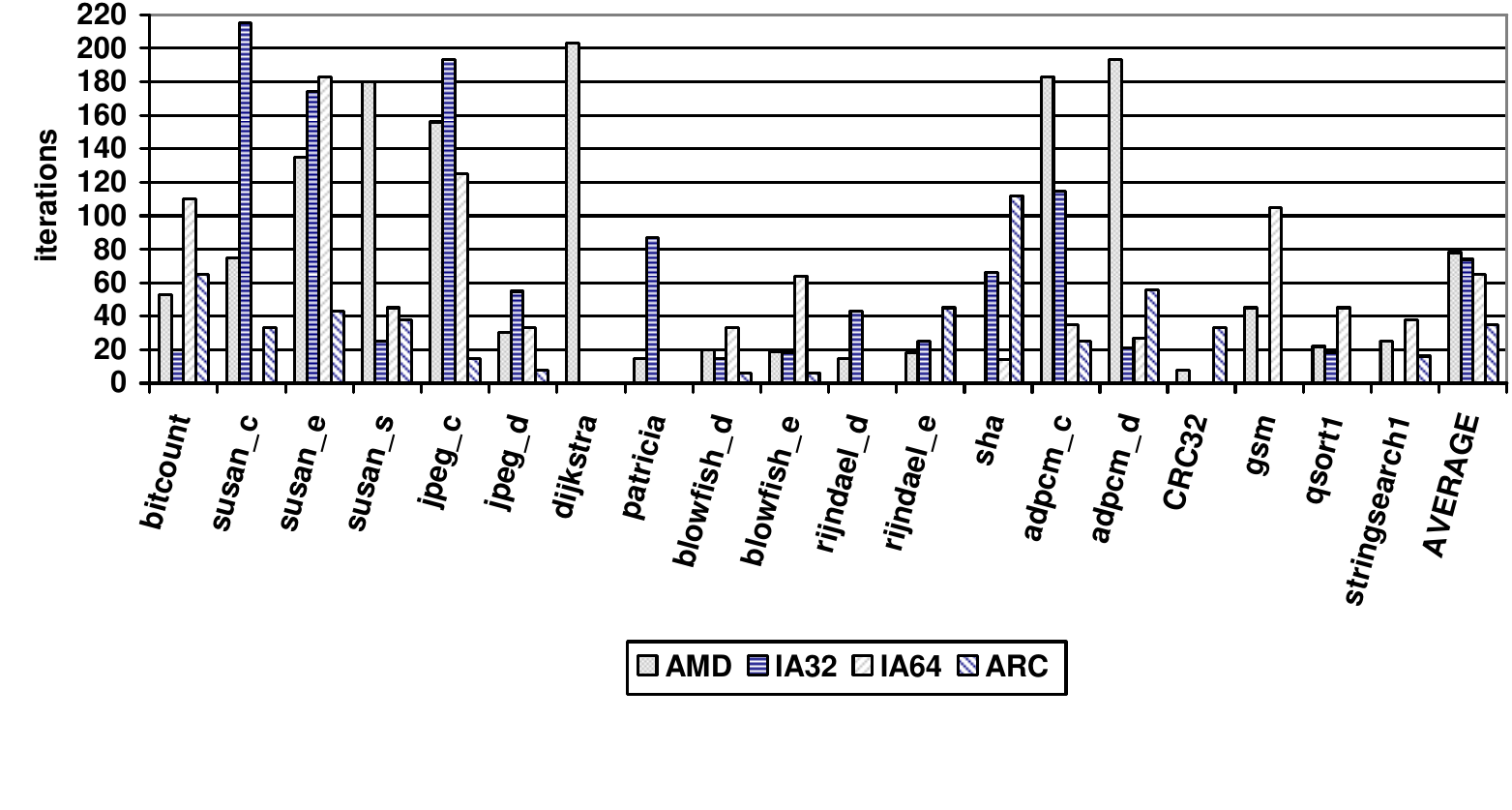}
  \caption{\it Number of iterations needed to obtain 95\% of the available speedup across cBench programs
using iterative compilation with uniform random distribution (500 iterations) on 4 distinct architectures.}
  \label{fig:speed}
\end{figure}

The problem of finding good combinations of optimizations or tuning
default compiler optimization levels becomes worse and more time consuming
when adding more transformations, optimizing for multiple datasets, 
architectures and their configurations,  adding more optimization
objectives such as reducing power consumption and architecture size,
improving fault-tolerance, enabling run-time parallelization and adaptation
for heterogeneous and reconfigurable multi-core systems and so on. 
Hence, in practice, it is not uncommon that compiler performs reasonably well 
only on a limited number of benchmarks and on a small range of a few relatively 
recent architectures but can underperform considerably on older or emerging architectures.

\begin{table*}[htb]
\begin{center}
\begin{tabular}{|p{450pt}|}
\hline
-O1 -falign-loops=10 -fpeephole2 -fschedule-insns -fschedule-insns2 -fno-tree-ccp -fno-tree-dominator-opts -funroll-loops \\ \hline
-O1 -fpeephole2 -fno-rename-registers -ftracer -fno-tree-dominator-opts -fno-tree-loop-optimize -funroll-all-loops \\ \hline
-O2 -finline-functions -fno-tree-dce -fno-tree-loop-im -funroll-all-loops \\ \hline
-O2 -fno-guess-branch-probability -fprefetch-loop-arrays -finline-functions -fno-tree-ter \\ \hline
-O2 -fno-tree-lrs \\ \hline
-O2 -fpeephole -fno-peephole2 -fno-regmove -fno-unswitch-loops \\ \hline
-O3 -finline-limit=1481 -falign-functions=64 -fno-crossjumping -fno-ivopts -fno-tree-dominator-opts -funroll-loops \\ \hline
-O3 -finline-limit=64 \\ \hline
-O3 -fno-tree-dominator-opts -funroll-loops \\ \hline
-O3 -frename-registers \\ \hline
-O3 -fsched-stalled-insns=19 -fschedule-insns -funroll-all-loops \\ \hline
-O3 -fschedule-insns -fno-tree-loop-optimize -fno-tree-lrs -fno-tree-ter -funroll-loops \\ \hline
-O3 -funroll-all-loops \\ \hline
-O3 -funroll-loops \\ \hline
\end{tabular}
\end{center}
\caption{Some of the profitable combinations of GCC 4.2.2 flags for multiple programs and benchmarks
including EEMBC, SPEC and cBench across distinct architectures
that improve both execution time and code size.}
\label{tab:best_flags}
\end{table*}

For example, Figure~\ref{fig:all_speedups} shows the best speedups achieved
on a range of popular and realistic benchmarks including EEMBC, SPEC and
cBench over the best GCC optimization level (-O3) after 500 iterations
using relatively recent mature GCC 4.2.2 (1.5 years old) and 2 mature
architectures: nearly 4 years old AMD Ahtlon64 3700+ and 2 years old
quad-core Intel Xeon 2800MHz. It clearly demonstrates that it is possible to
considerably outperform even mature GCC with the highest default
optimization level using random iterative search. Moreover, it also 
shows that achievable speedups are architecture dependent and vary
considerably for each program ranging from a few percent to nearly 
two times improvements.

Figure~\ref{fig:speed} shows that it may take around 70 iterations
on average before reaching 95\% of the speedup available after 500 iterations for
cBench/MiBench benchmark and is heavily dependent on programs and
architectures. Such a large number of iterations is needed due to continuously
increasing number of aggressive optimizations available in the compiler that
can both considerably increase or degrade performance or change code size
making it more time consuming and non-trivial to find a profitable combination 
of optimizations in each given case. For example, Table~\ref{tab:best_flags} shows 
such non trivial combinations of optimizations that improve both execution time
and code size found after uniform random iterative compilation~\footnote{
After empirical iterative compilation with random uniform distribution 
we obtain profitable combinations of optimizations that may consist of 50 flags on average.
However, in practice, only several of these flags influence performance or code size.
Hence, we use CCC~\emph{ccc-run-glob-flags-one-off-rnd} plugin to prune found combinations
and leave only influential optimizations to improve performance analysis and optimization predictions.
} across all benchmarks and architectures for GCC 4.2.2.
One may notice that found combinations of profitable compiler optimizations 
also often reduce compilation time since some combinations of optimizations 
require only a minimal optimization level -O1 together with several profitable flags. 
Some combinations can reduce compilation time by 70\% which can be critical when compiling
large-scale applications and OS. All this empirical optimization information is 
now available in COD~\cite{ctuning_repository} for further analysis and improvement of a compiler design.

\begin{figure}[htb]
 \centering
 \includegraphics[width=6.5in]{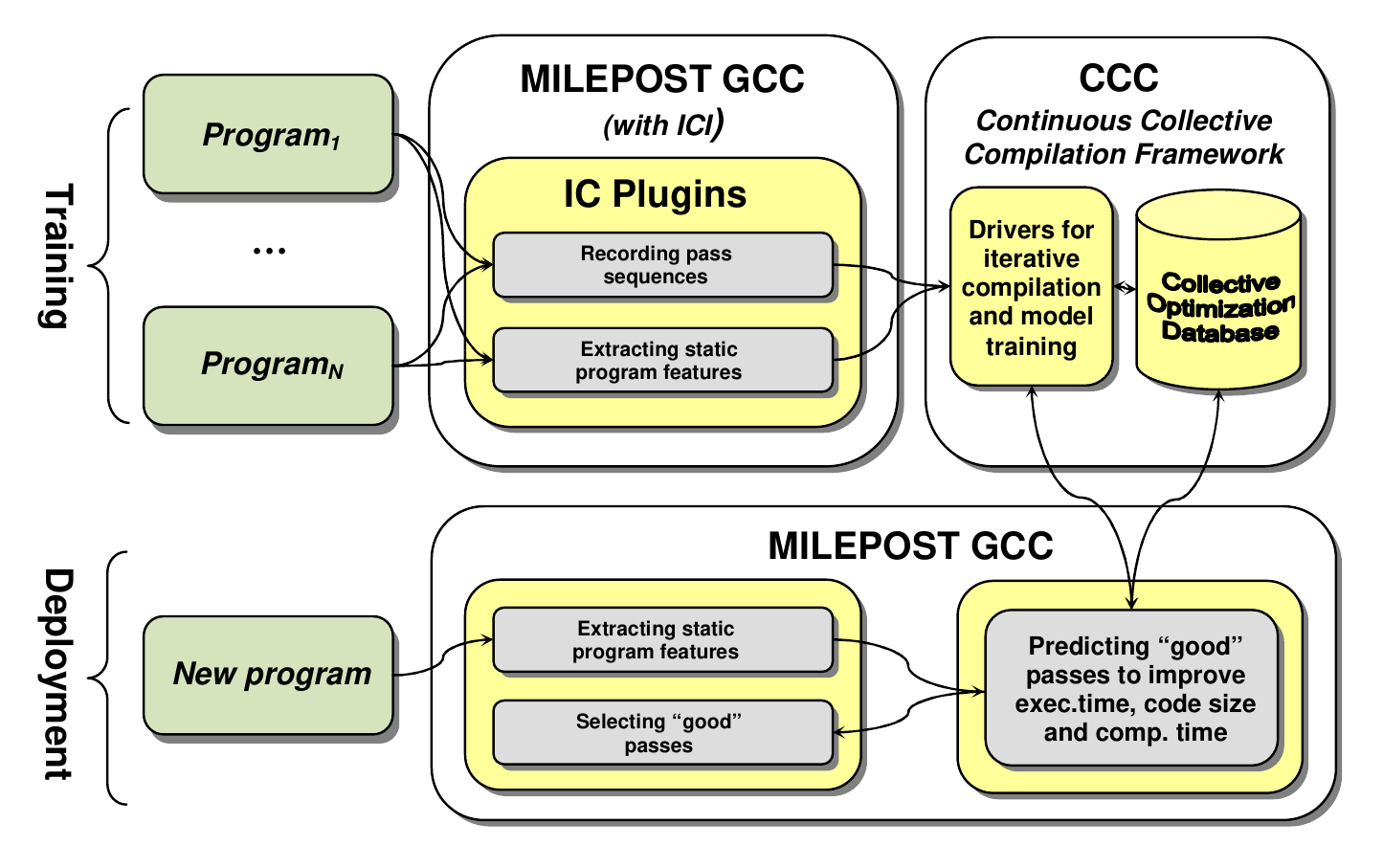}
 \caption{Original MILEPOST framework connected with cTuning infrastructure
 to substitute default compiler optimization heuristic  with an optimization
 prediction plugin based on machine learning. It includes MILEPOST GCC with
 Interactive Compilation Interface (ICI) and program features extractor, and
 CCC~Framework to train ML model and share optimization data in COD.}
 \label{fig:ml}
\end{figure}

We expect that optimization spaces will increase dramatically after we
provide support for a fine-grain optimization selection and tuning in GCC
using ICI~\cite{gsoc2009}. In such situation, we hope that cTuning
technology will considerably simplify and automate the exploration of large
optimization spaces with "one button" approach when a user just controls and
balances several optimization criteria.

\begin{figure}[htb]
 \centering
 \includegraphics[width=5in]{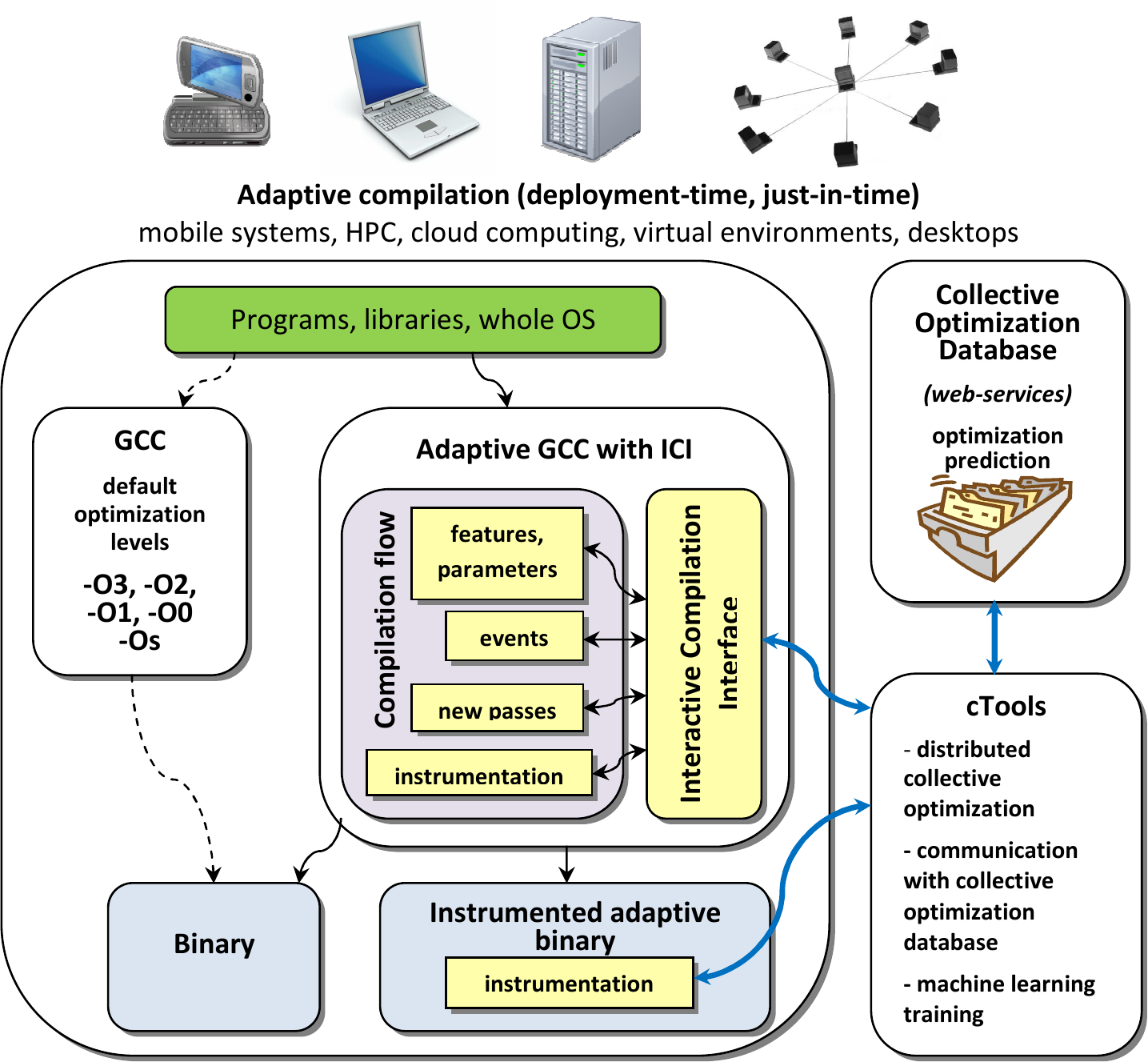}
 \caption{Using cTuning optimization prediction web services to substitute default compiler optimization levels and 
predict good optimizations for a given program and a given architecture on the fly based on its structure or
or dynamic behavior (program static and dynamic features) and continuously retrained predictive models based on collective
optimization concept~\cite{FT2009}.}
 \label{fig:ctuning_prediction_service}
\end{figure}

\subsection{MILEPOST GCC and optimization prediction web services based on machine learning}

\begin{figure}[htb]
 \includegraphics[width=5in]{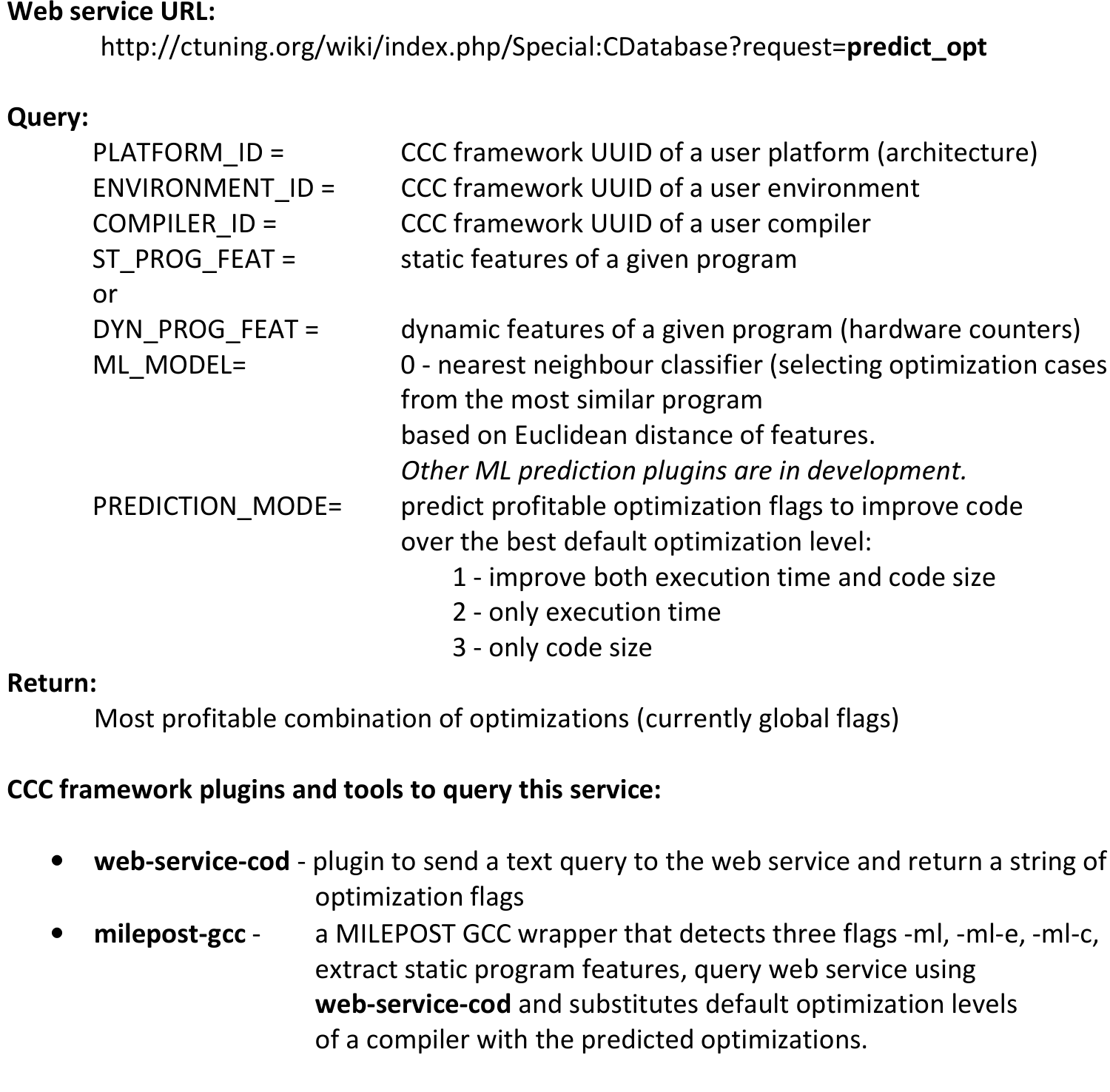}
 \caption{cTuning optimization prediction web service: URL, message format and tools.}
 \label{fig:ctuning_prediction_service_query}
\end{figure}

Default optimization levels in compilers are normally aimed to deliver good
average performance across several benchmarks and architectures relatively
quickly. However, it may not be good enough or even acceptable for many
applications including performance critical programs such as real-time
video/audio processing systems, for example. That is clearly demonstrated in
Figure~\ref{fig:all_speedups} where we show the improvements in execution time
for multiple popular programs and several architectures of nearly 3 times over 
the best default optimization level of GCC using random feedback-directed 
compilation after 500 iterations.

\begin{figure}[tb]
 \centering
 \includegraphics[width=6.5in]{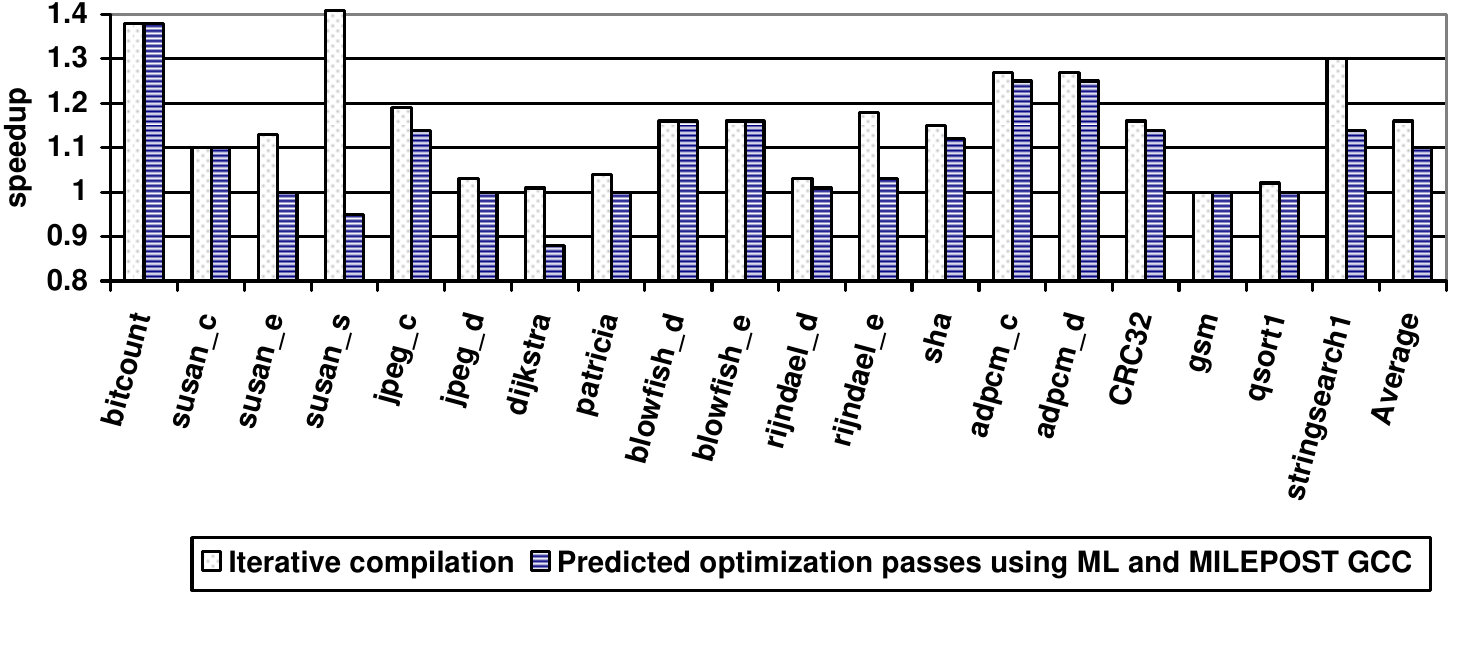}
 \caption{Speedups achieved when using iterative compilation on ARC with random search strategy 
(500 iterations; 50\% probability to select each optimization;) and when predicting best optimizations
using \emph{probabilistic ML model} based on program features described in~\cite{ABCP06}}
 \label{fig:fig_results}
\end{figure}

In~\cite{FMTP2008} we introduced our machine learning based research
compiler (MILEPOST GCC) and integrated it with the cTuning infrastructure
during the MILEPOST project~\cite{milepost} to address the above problem by
substituting default optimization levels of GCC with a predictive
optimization heuristic plugin that suggests good optimizations for a given
program and a given architecture.

Such framework functions in two distinct phases, in accordance with typical
machine learning practice: training and deployment as shown in Figure~\ref{fig:ml}. 

During the training phase we gather information about the structure of
programs (static program features) and record how they behave when compiled
under different optimization settings (execution time or other dynamic
program features such as hardware counters, for example) using CCC
framework. This information is recorded in COD and used to correlate program
features with optimizations, building a machine learning model that predicts
a good combination of optimizations.

In order to train a useful model a large number of compilations and
executions as training examples are needed. These training examples are now
continuously gathered from multiple users in COD together with program
features extracted by MILEPOST GCC.

Once sufficient training data is gathered, different machine learning models
are created based on (probabilistic and decision trees approaches among
others)~\cite{FMTP2008,ABCP06,CFAP2007}, for example. We provided such
models trained for a given architecture as multiple architecture-dependent
optimization prediction plugins for MILEPOST GCC. When encountering a new
program, MILEPOST GCC extracts program features and passes them to the
ML plugin which determines which optimizations to apply. 

When more optimization data is available (through collective
optimization~\cite{FT2009}, for example) or when some new transformations
are added to a compiler, we need to retrain our models for all architectures
and provide new predictive optimization plugins for downloading simplifying
and modularizing a compiler itself.

Naturally, since all filtered static and dynamic optimization data is now continuously gathered in COD,
we can also continuously update optimization prediction models for different architectures at cTuning website. 
Hence, we decided to create a continuously updated online optimization prediction web-service as shown in 
Figure~\ref{fig:ctuning_prediction_service}. It is possible to submit a query to this web service
as shown in Figure~\ref{fig:ctuning_prediction_service_query} providing information about architecture, 
environment, compiler, static or dynamic program features and selecting a required machine learning model
and an optimization criteria, i.e. improving either execution time or code size or both over best default
optimization level of a compiler. At the moment, this web service returns the most profitable combination 
of global compiler optimizations (flags) to improve a given program. This service can be tested online
at~\cite{ctuning_cpredict} or using some plugins from CCC framework to automate prediction.

Such optimization prediction web service opens up many optimization and
research possibilities: We plan to test it to improve the whole OS
optimization (Gentoo-like Linux, for example), improve adaptation of
downloadable applications for a given architecture (Android and Moblin
mobile systems, cloud computing and virtual environments, for example),
just-in-time optimizations for heterogeneous reconfigurable architectures
based on program features and run-time behavior among others. Of course, we
need to minimize Internet traffic and queries. Hence, we will need to
develop an adaptive local optimization proxy to keep associations between
local program features and optimizations for a given architecture while
occasionally updating them using global cTuning web services. We leave it
for the future work.

As a practical example of the usage of our service, we trained our
prediction model for an ARC725D reconfigurable processor using MILEPOST GCC
and cBench with 500 iterations and 50\% probability of selecting each
compiler flag for an individual combination of optimizations at each iteration.
Figure~\ref{fig:fig_results} compares the speedups achieved after training
(our execution time upper bound) and after one-shot optimization prediction
(as described in detail in~\cite{ABCP06}).It demonstrates that except a few
pathological cases we can automatically improve original production ARC GCC
by around 11\% on average using cTuning infrastructure.

%%%%%%%%%%%%%%%%%%%%%%%%%%%%%%%%%%%%%%%%%%%%%%%%%%%%%%%%%%%%%%%%%%%%%%%%%
\section{Conclusions and Future Work}
\label{sec:final}

In this paper we presented our long-term collective tuning initiative to
automate, distribute, simplify and systematize program optimization,
compiler design and architecture tuning using empirical, statistical and
machine learning techniques. It is based on sharing of empirical
optimization experience from multiple users in the Collective Optimization
Database, using common collaborative R\&D tools and plugin-enabled production
quality compilers with open APIs and providing web services to
predict profitable optimizations based on program features.

We believe that cTuning technology opens up many research, development and
optimization opportunities. It can already help to speed up existing
underperforming computing systems ranging from small embedded architectures
to high-performance servers automatically. It can be used for a more realistic
statistical performance analysis and benchmarking of computing systems. It
can enable statically-compiled self-tuning adaptive binaries and libraries.
Moreover, we believe that cTuning initiative can improve the quality and
reproducibility of academic and industrial IT research. Hence, we decided to
move all our developments to public
domain~\cite{ctuning,ctuning_mailing_lists} to enable collaborative
community-based developments and boost research. We hope that using common
cTuning tools and optimization repository can help to validate research
ideas and move them back to the community much faster.

We promote top-down optimization approach starting from global and 
coarse-grain optimizations and gradually supporting more fine-grain
optimizations to avoid solving local optimization problems without understanding 
the global optimization problem first. Within Google Summer of Code'2009 program~\cite{gsoc2009},
we plan to enable automatic and transparent collective program optimization and run-time
adaptation based on~\cite{FT2009} providing support for fine grain program optimizations
and reordering, generic function cloning and program instrumentation in GCC using ICI.
We will also need to provide formal validation of code correctness during transparent
collective optimization. We plan to combine CCC framework with architectural
simulators to enable systematic software/hardware co-optimization. 
We are also extending UNIDAPT framework to improve automatic profile-driven statistical
parallelization and scheduling for heterogeneous multicore architectures~\cite{JGVP2009,LCWP2009,TWFP2009,LFF2007}
using run-time monitoring of data dependencies and automatic data partitioning and scheduling
based on static and dynamic program/dataset features combined with machine learning and
statistical techniques.

We are interested to validate our approach in realistic environments and
help better utilize available computing systems by improving whole OS
optimizations, adapting mobile applications for Android and Moblin on the
fly, optimizing programs for grid and cloud computing or virtual
environments, etc. Finally, we plan to provide academic research plugins and
online services for optimization data analysis and visualization.

To some extent, cTuning concept is similar to biological self-tuning environments
since all available programs and architectures can be optimized slightly differently
continuously favoring the most profitable optimizations and designs over time.
Hence, we would like to use cTuning knowledge to start investigating 
completely new programming paradigms and architectural designs 
to enable development of the future self-tuning and self-organizing 
computing systems.

%%%%%%%%%%%%%%%%%%%%%%%%%%%%%%%%%%%%%%%%%%%%%%%%%%%%%%%%%%%%%%%%%%%%%%%%%
\section{Acknowledgments}

cTuning development has been partially supported by the MILEPOST
project~\cite{milepost}, HiPEAC network of excellence~\cite{hipeac} and
Google Summer of Code'2009 program~\cite{gsoc2009}. I would like to thank
Olivier Temam for interesting discussions about Collective Optimization
concept and Michael O'Boyle for machine learning discussions; Zbigniew
Chamski for his help with the development of the new ICI; Mircea Namolaru
for the development of the static program feature extractor for GCC with
ICI; Erven Rohou for his help with cBench developments; Yang Chen, Liang
Peng, Yuanjie Huang, Chengyong Wu for cTools discussions, feedback and help
with UNIDAPT framework extensions. I would also like to thank Abdul Wahid
Memon and Menjato Rakototsimba for thorough testing of the Continuous Collective
Compilation Framework and Collective Optimization Database. Finally, I would
like to thank cTuning and GCC communities~\cite{ctuning_mailing_lists,
gcc_org} for very useful feedback, discussions and help with software
developments.

\bibliographystyle{abbrv}
\bibliography{fursin}

\end{document}